\documentclass[12pt,final,english]{iopart}
\usepackage{xcolor}
\usepackage{graphicx}
\usepackage{amssymb}
\usepackage{setspace}
\usepackage{verbatim}
\usepackage{listings}
\usepackage[backend=biber, style=numeric, sorting=none]{biblatex}

\begin{document}


\title{Disambiguation of magnetic sensors in ITER}

\author{
M. J. Hole$^{1,2}$, 
S. D. Pinches$^3$, 
G. Vayakis$^3$,
S. C. McIntosh$^3$,
O. Hoenen$^3$, 
P. Abreu$^3$}

\address{$^1$
Mathematical Sciences Institute and School of Computing, the Australian National University, Canberra ACT 2601, Australia}

\address{$^2$
Australian Nuclear Science and Technology Organisation, Locked Bag 2001, Kirrawee DC, NSW 2232, Australia}

\address{$^3$
ITER Organization, Route de Vinon-sur-Verdon, CS 90 046, 13067 St. Paul Lez Durance Cedex, France}


\begin{abstract}

ITER will possess approximately 500 magnetic sensors (mainly measuring poloidal flux) distributed across the first wall. The coils are at known locations but the matching signals not necessarily known. There may also be mistakes in the wiring of the coil polarity. The existing strategy to disambiguate coils uses combinatoric programming of poloidal field coil waveforms of up to 48 discharges of plasma-less operation. An alternate strategy explored in this work is the energisation of a combination of both poloidal and toroidally asymmetric active coils, and Biot-Savart computation of the field solution from all active coils at the sensor coils. A direct brute force permutation of all $N$ coil combinations scales as $O(2^N N!) $ which is intractable for $N>10$.  The mathematically formulated optimisation problem was analysed using AI-assisted coding tools, which identified the problem structure as a signed assignment problem and suggested a Hungarian-algorithm-based optimisation strategy, which scales as $O(N^3)$.  This search algorithm, when embedded into the magnetic-diagnostic identification problem, was able to disambiguate randomly connected and polarised coils in a regularly spaced array in the ITER first wall (where the the coils are located) down to a signal-to-noise ratio of 50. The computation took 2 seconds. Reconstruction of the actual coil positions in the ITER first wall was achieved high confidence, $C>0.97$.  Reconstruction of the second wall poloidal flux coils, which comprised multiple arrays at near constant $\phi$ (each of which is regularly spaced in $\theta$) had a much lower confidence, of $C > 0.15$.  By adding the active poloidal field coils to the combined cost function, the confidence increased to $C > 0.59$. This provides the opportunity to reduce the commissioning time of ITER, and is a strategy that could be tested on other toroidal magnetic confinement devices.

\end{abstract}

\submitto{\PPCF}
\maketitle

\section{Introduction}


Magnetic diagnostics play a critical role in the operation, control, and protection of toroidal magnetically confined fusion plasmas, especially tokamaks.  \cite{Pustovitov_2001} Existing work has focused on design \cite{Hutchinson_2002}, calibration \cite{Heeter_2000, Appel_Hole_2005}, alignment and effective area determination \cite{Abate_2022}, integration drift \cite{Moret_1998}, and magnetic reconstruction \cite{Lao_1985,Lao_2005}.  However, comparatively little attention has been paid to the combinatorial ambiguity problem associated with sensor polarity, ordering, or wiring identification during installation and commissioning. In large systems containing many sensors, this naturally leads to a signed assignment problem, motivating the optimisation framework developed in this work.

In ITER, an extensive network of approximately 500 magnetic sensors —predominantly poloidal flux loops and field probes mounted on the inner and outer vacuum vessel walls— will provide crucial information on the equilibrium, stability, and dynamics of the plasma. Although the geometric locations of these sensors are precisely defined at the design stage, their electrical mapping to data acquisition channels, including the polarity of their connections, will not be resolved in ITER during assembly.  \cite{ITER_55.A0_2024} (Even in experiments where great care is taken, the wiring and polarity cannot be assumed to be error‑free after installation.) Disambiguation is essential for magnetic reconstruction, real time control, and machine operation. 


Existing disambiguation strategies \cite{ITER_55.A0_2024} rely on energising combinations of axisymmetric poloidal field (PF) coils in a sequence of plasma‑less commissioning discharges. By comparing the resulting magnetic signals against predictions from calibrated field models, one attempts to infer the correct mapping between sensors and data channels.  However, for the full ITER system this approach may require on the order of several dozen discharges, because only a limited number of sufficiently distinguishable field configurations can be produced using axisymmetric coils alone. Moreover, as the number of sensors grows, the combinatorial complexity of brute‑force matching between predicted and measured signals increases factorially, making a direct search intractable even for modest sensor counts.

In this work, we introduce a new methodology that dramatically accelerates sensor identification by leveraging asymmetric active field coils—specifically, ITER’s resonant magnetic perturbation (edge control) coils and correction coils—in combination with a computationally efficient optimisation algorithm. By using these toroidally localised and spatially distinct field sources, we generate magnetic signatures with greater discriminatory power across the vessel surface. We compute the corresponding magnetic fields using Biot–Savart integration and compare them with synthetic or measured sensor data that may contain both permutation and sign errors.

A key innovation of the study is the application of the Hungarian algorithm \cite{Kuhn_1955} to this “signed assignment” problem. Whereas a naïve brute‑force approach scales as $O(2^N N!)$ and becomes infeasible for $N >10$ , the Hungarian method scales as $O(N^3)$, enabling rapid identification even for ITER‑scale diagnostic sets. We demonstrate that the technique can correctly recover sensor indexing and polarity at signal‑to‑noise ratios as low as 50–80, with typical execution times of order seconds on standard hardware. Importantly, we also introduce a confidence metric that quantifies the robustness of the inferred mapping and highlights potential degeneracies.

The development of our statistical confidence measures and the resolution of degenerate solutions builds on earlier work in diagnostic optimisation. 
Previous work by Hole and Appel developed Fourier-SVD techniques for synchronous decomposition of magnetic perturbations in toroidal plasmas, including statistical confidence measures and resolution of degenerate mode structures. \cite{Hole_2007b} Subsequent work applied optimisation principles to the design and placement of high-frequency Mirnov arrays for maximal mode-number resolving power in MAST. \cite{Hole_OMAHA_2009} The present work extends these ideas to the combinatorial problem of sensor identity and polarity reconstruction, formulating the task as a signed assignment optimisation problem.

The strategy presented here has the potential to significantly reduce the commissioning time of ITER’s magnetic diagnostics and can be readily adapted to other tokamak systems. It offers a scalable, generalisable, and computationally efficient approach to ensuring that magnetic measurements—which underpin plasma control and machine protection—are correctly interpreted from the outset of device operation.

The remainder of this section describes the passive coils, Biot-Savart toolbox to compute solutions and the measurement plane.   Section \S \ref{sec:ITER_coils} describes the active coils in ITER and the solutions on the measurement plane, while \S \ref{sec:strategy} introduces the search strategy and \S \ref{sec:ITER_app} applies the strategy to ITER. Finally, \S \ref{sec:conclusions} concludes the work.

\subsection{Field Modelling}

The commissioning discharges considered here evolve sufficiently slowly that magnetostatic Biot–Savart solutions are adequate to describe the vacuum magnetic field structure.  We have computed solutions to Biot-Savart equation 
\begin{equation}
    \mathbf{B}(r) = \frac{\mu_0}{4 \pi} \int_C \frac{I \mathbf{dl} \times (\mathbf{r} - \mathbf{r'})}{|\mathbf{r} - \mathbf{r'}|^3}
\end{equation}
for arbitrary conducting filaments using the BSmag Toolbox \cite{BSmag_2015}. Here, $I$ is the current in the filament, $\mathbf{dl}$ the vector representing the infinitessimal length of the current filament in the direction of the current, $\mathbf{r'}$ the position of the current filament, and $\mathbf{r}$ position.  The BSmag toolbox is a MATLAB-based package that computes the magnetic field using direct numerical integration of the Biot–Savart line integral for filamentary currents. It can calculate both the field $\mathbf{B}$ and its spatial derivatives from arbitrary conducting paths. Filament paths are discretised into piecewise linear segments $\gamma = \mathbf{dl}$ with $d\gamma$ the maximum filament discretisation length. 


\subsection{Magnetic Sensors}
\begin{figure}
 \centering
       (a)\includegraphics[width=0.5\textwidth]{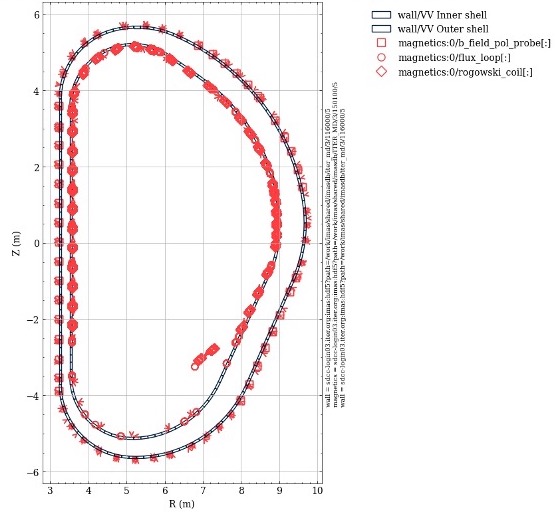}
       (b)\includegraphics[width=0.3\textwidth]{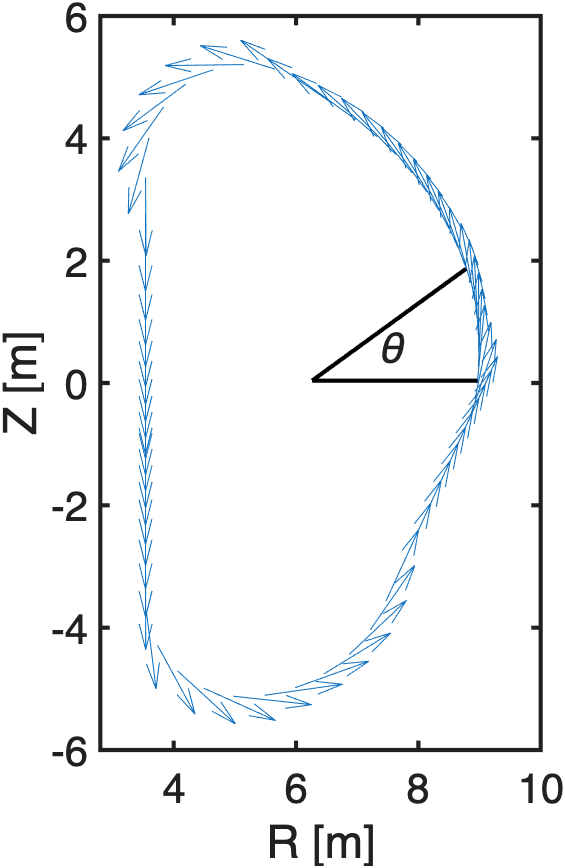}
 \caption{ (a) Poloidal cross section of ITER showing inner and outer vessel walls, location of poloidal field probes (square), flux loops (circle) and Rogowski coils (diamond). (b) The ITER inner wall, parameterisation, geometric angle $\theta$ with axis located at the geometric centre $(6.2523, 0.0767)$, and the tangent vector field.}\label{fig1:ITER_sensors}
\end{figure}

The equilibrium magnetic sensors of ITER are largely poloidal magnetic flux probes and are located on the inner or outer vacuum vessel.  Figure \ref{fig1:ITER_sensors}(a) shows a cross-section of the wall with magnetic diagnostics insert.   The poloidal probes are oriented tangent to the wall, and therefore measure the tangential magnetic field component.  We have parameterised the wall by $R_w = f_R(\theta), Z_w = f_Z (\theta)$ where the geometric angle $\theta$ rotates clockwise from the outboard midplane, as shown in Fig. \ref{fig1:ITER_sensors}(b), and $\theta=0$ is located at the geometric centre.  The tangent vector field $\bf{\tau}$ is hence given by 
\begin{equation}
\tau = \frac{f_R'(\theta)}{\sqrt{f_R'(\theta)^2 + f_Z'(\theta)^2}} \hat{\bf{R}} + 
\frac{f_Z'(\theta)}{\sqrt{f_R'(\theta)^2 + f_Z'(\theta)^2}} \hat{\bf{Z}}   \label{eq:wall_tang}
\end{equation}
Figure \ref{fig1:ITER_sensors}(b) shows the tangent vector field for the inner wall.  The full space of solutions is thus given in $(\theta,\phi)$ space, with $\phi$ the geometric toroidal angle.  The variation of $\mathbf{B} \cdot \tau$ with poloidal and toroidal position provides the spatial structure exploited for sensor disambiguation.


\section{ITER Active Field Coils}\label{sec:ITER_coils}
ITER comprises axis-symmetric active field coils (poloidal field coils PF1-PF6, and fast control coils VS1 and VS2), as well as axis asymmetric field coils (resonant magnetic perturbation or edge control coils ECC1-ECC27), and correction coils CC1-CC9). 

\subsection{Axis-symmetric Coils}

\begin{figure}
 \centering
       (a)\includegraphics[width=0.3\textwidth]{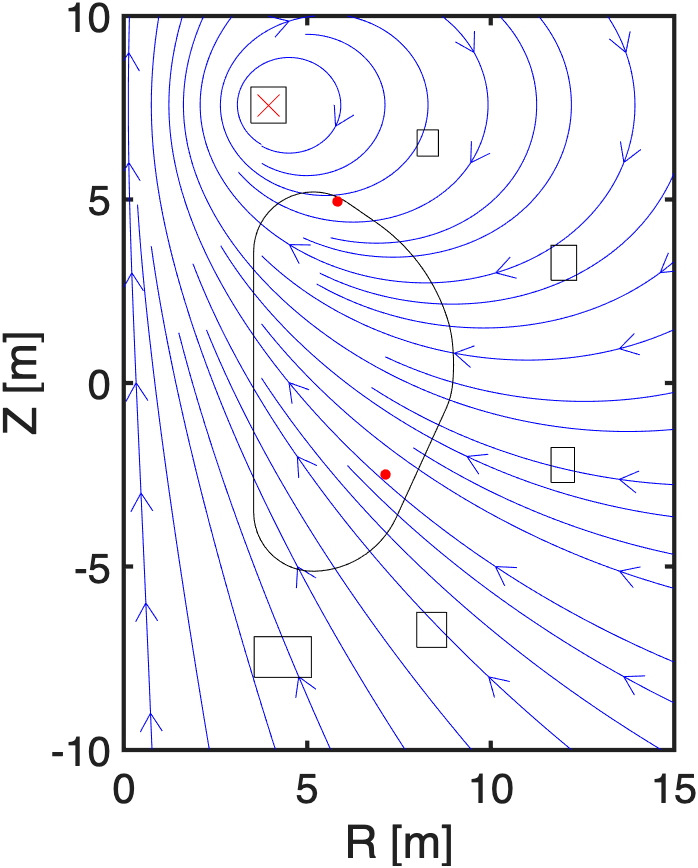}
       (b)\includegraphics[width=0.4\textwidth]{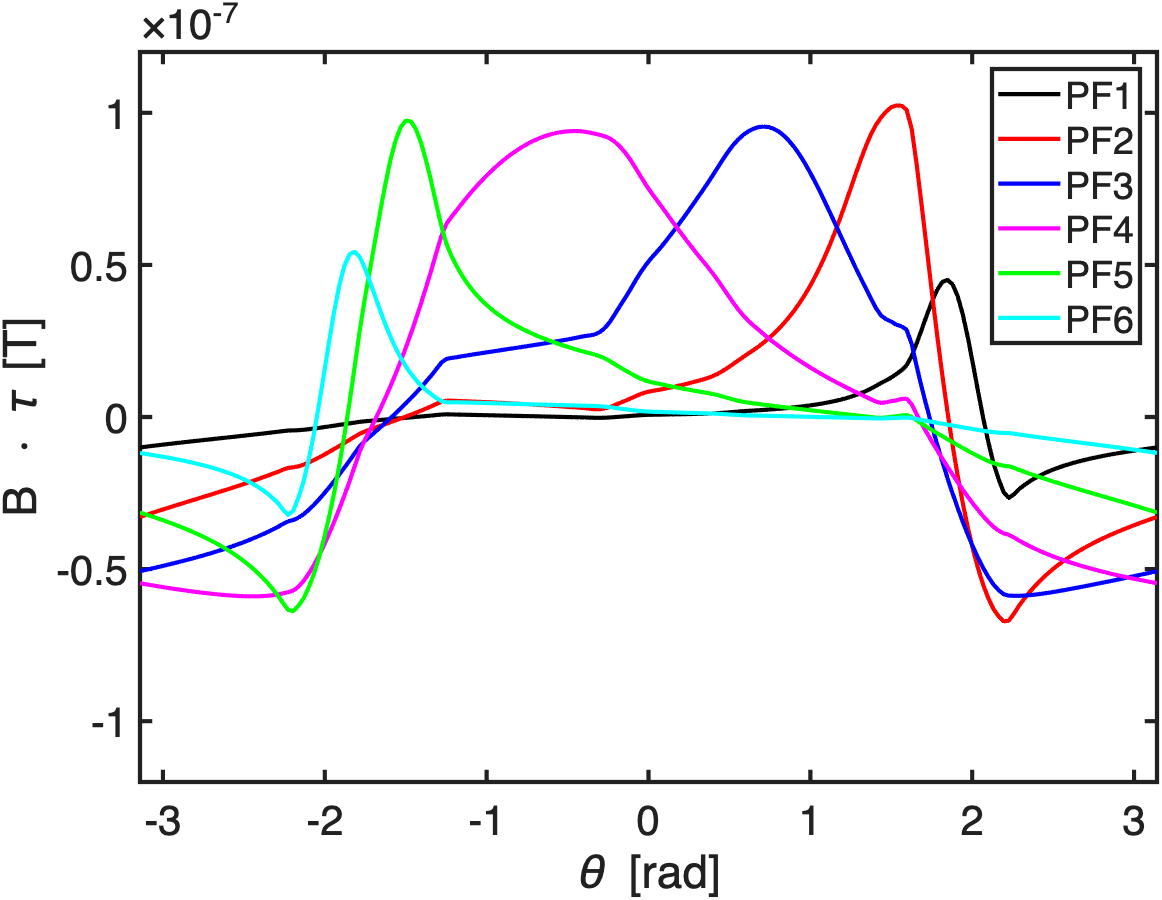}
 \caption{ (a) Poloidal cross section of ITER showing the inner vessel walls, casing of the poloidal field coils (square), in-vessel control coils (red circles interior to wall). Inset shown in blue are stream lines of the poloidal magnetic field computing by PF1, centred at the red cross at $R=3.95$~m and $Z=7.57$~m. (b) Computation of $\mathbf{B} \cdot \mathbf{\tau}$ for PF1-PF6 as a function of $\theta$.}\label{fig:ITER_pf_coils}
\end{figure}

Figure \ref{fig:ITER_pf_coils}(a) is a poloidal cross-section of ITER showing PF1-PF6 (casings) as well as VS1, VS2.  Overplotted are streamlines of the field for PF1 using the BSmag tool box, and treating the coil in the filament limit. We have then computed $\mathbf{B} \cdot \mathbf{\tau}$ for the different poloidal field coil activations for a current of 1A. The outermost poloidal field coils PF1 and PF6 produced extended poloidal regions for which $\mathbf{B} \dot \tau$ is small.



\subsection{Edge Control Coils}

\begin{figure}
 \centering
       (a)\includegraphics[width=0.4\textwidth]{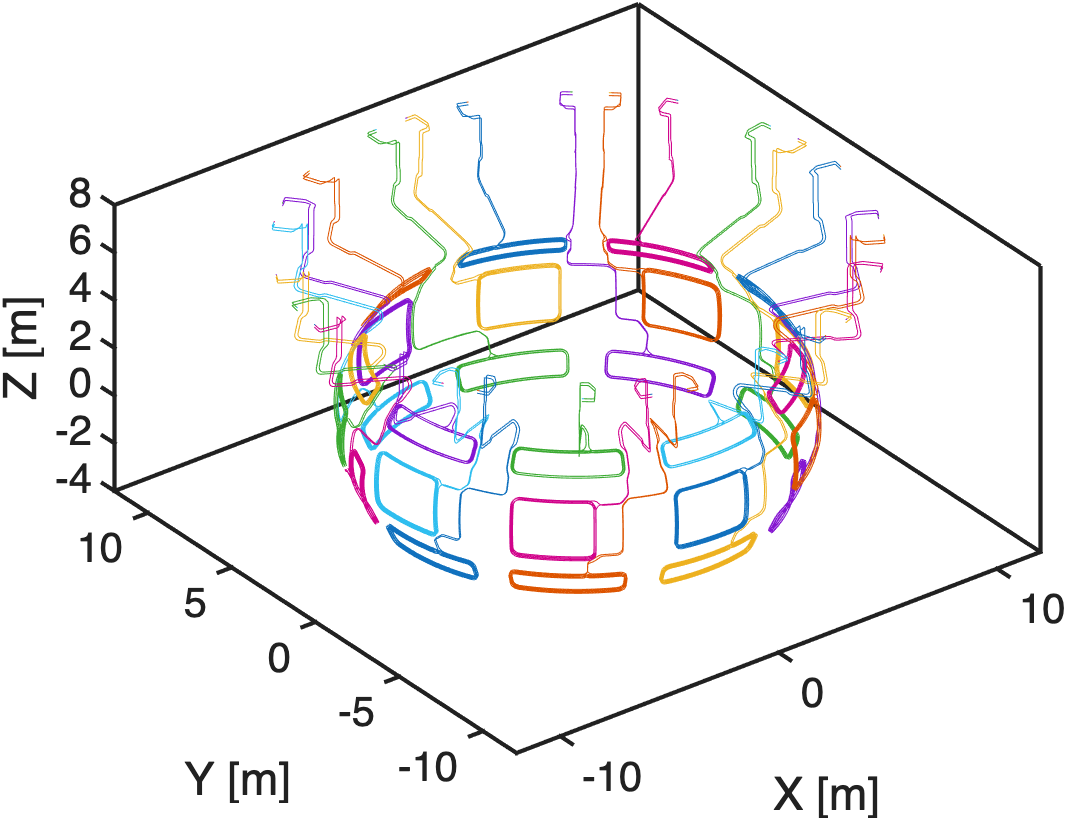}
       (b)\includegraphics[width=0.4\textwidth]{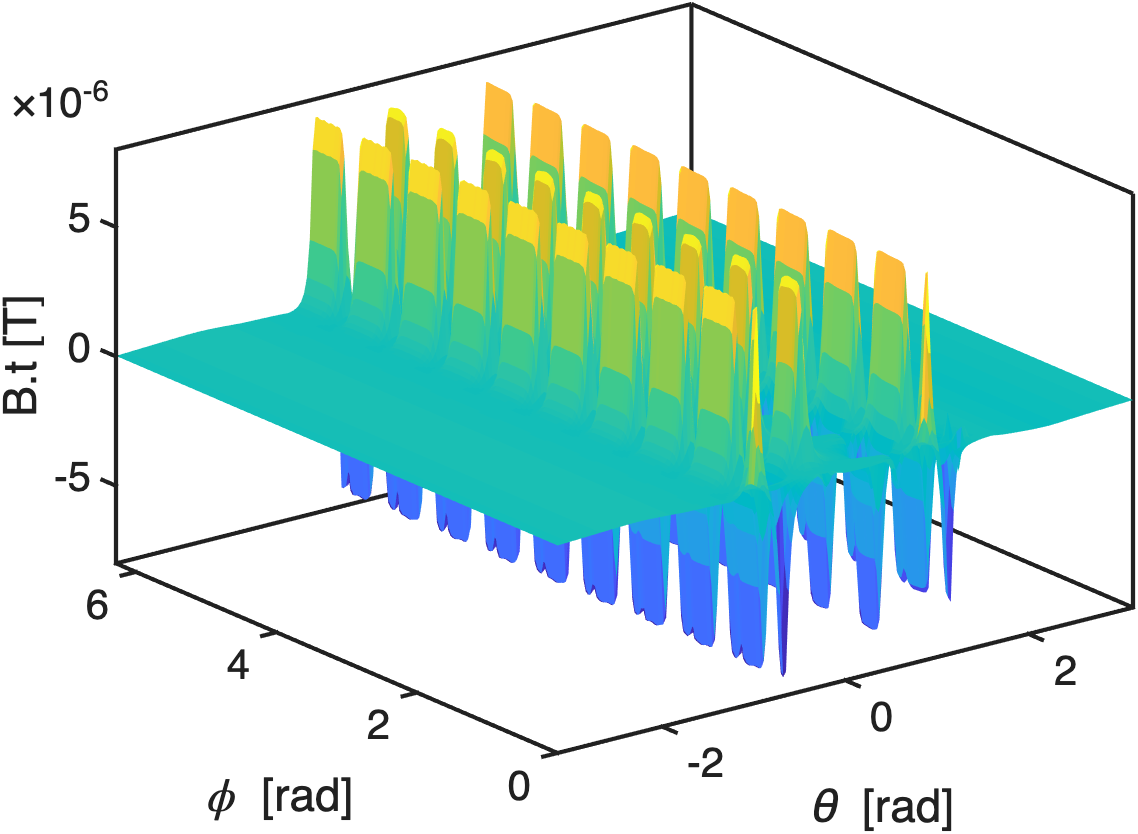}
       (c)\includegraphics[width=0.4\textwidth]{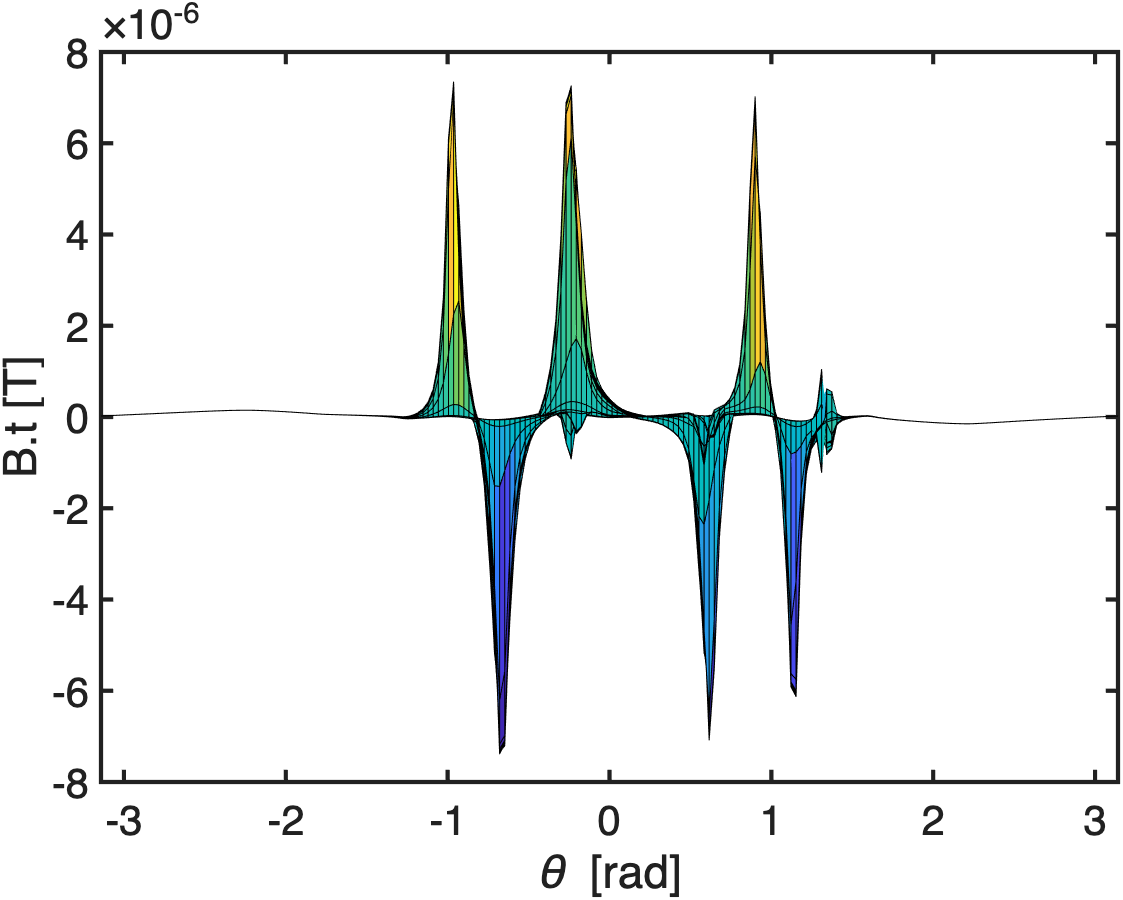}
 \caption{ (a) ITER edge control coils ECC1-ECC27 with feedthrough, (b) computed $\mathbf{B} \cdot \mathbf{\tau}$ plotted over unwrapped $(\phi, \theta)$ space and (c) $\mathbf{B} \cdot \mathbf{\tau}$ plotted with $\mathbf{\phi}$ into page.
 }\label{fig:ITER_rmp_coils}
\end{figure}

Figure \ref{fig:ITER_rmp_coils}(a) shows the Edge Control Coils (ECC) including feedthroughs, together with the computed $\mathbf{B} \cdot \mathbf{\tau}$ for 1A in $(\phi, \theta)$ space and with $\mathbf{\phi}$ into page.  The field is substantially more localised to $(\phi, \theta)$ of each coil.  In the poloidal direction this is the outboard from $-\pi/2 <\theta < \pi/2 $.  



\subsection{Correction Coils}

\begin{figure}
 \centering
       (a)\includegraphics[width=0.4\textwidth]{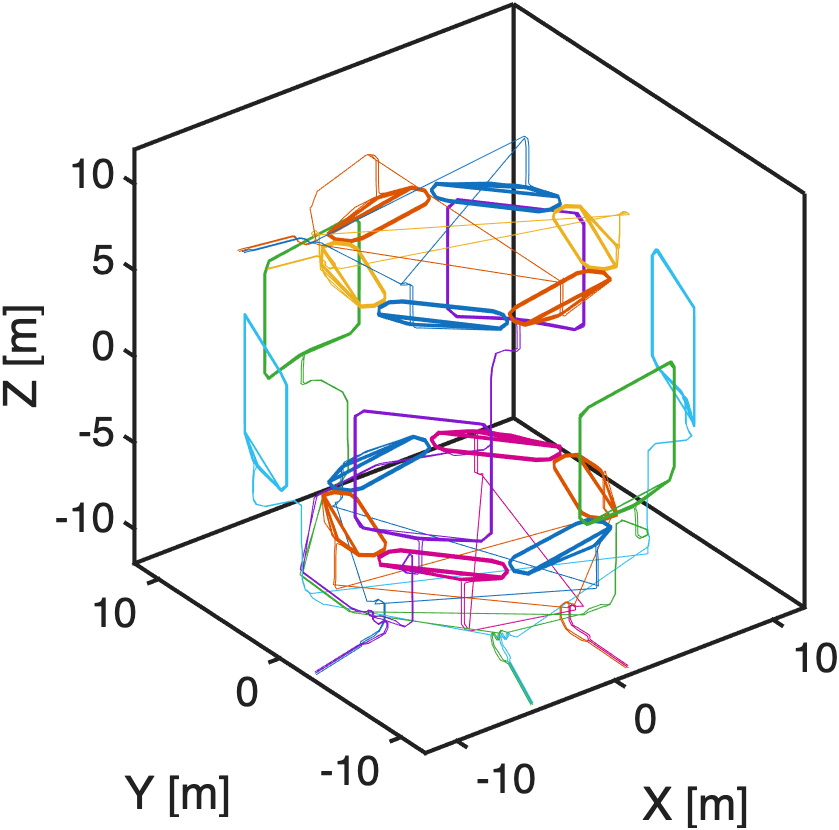}
       (b)\includegraphics[width=0.4\textwidth]{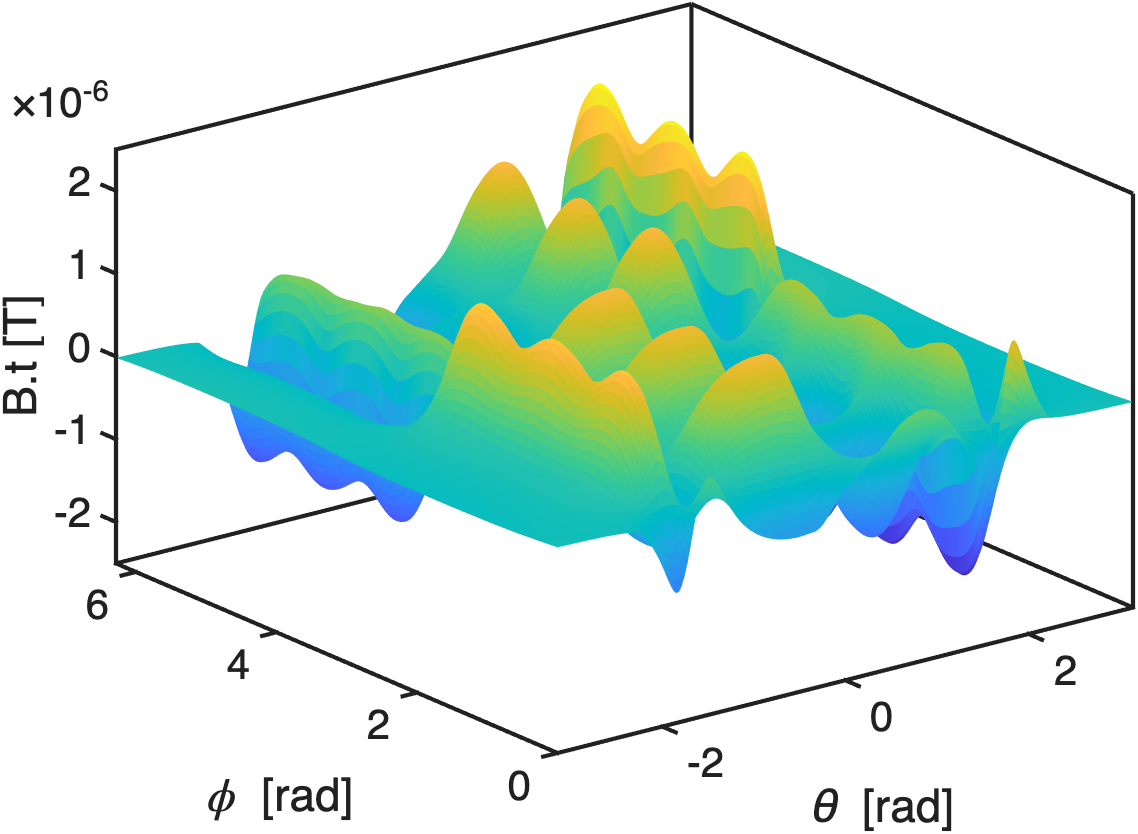}
       (c)\includegraphics[width=0.4\textwidth]{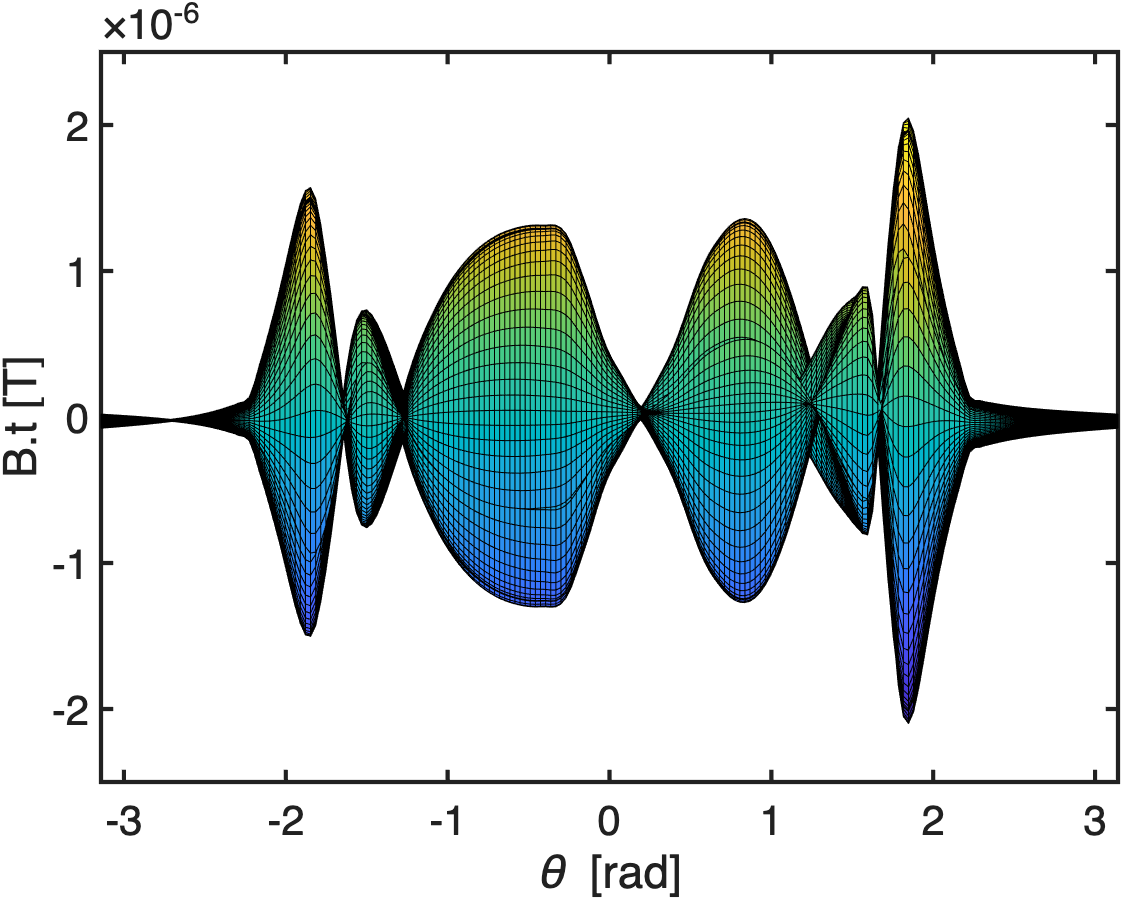}
 \caption{ (a) ITER edge control coils CC1-CC9 with feedthrough, (b) computed $\mathbf{B} \cdot \mathbf{\tau}$ plotted over unwrapped $(\phi, \theta)$ space and (c) $\mathbf{B} \cdot \mathbf{\tau}$ plotted with $\mathbf{\phi}$ into page.
 }\label{fig:ITER_cc_coils}
\end{figure}

Figure \ref{fig:ITER_cc_coils}(a) shows the Correction Coils (CC) including feedthroughs, together with the computed $\mathbf{B} \cdot \mathbf{\tau}$ for 1A in $(\phi, \theta)$ space and with $\mathbf{\phi}$ into page.  As expected, the field pattern is significantly larger across the plasma volume and the wall.


\section{Search Strategy}\label{sec:strategy}

A succinct statement of the optimisation strategy is as follows.   Suppose there are $N= N_\phi N_\theta$ magnetic sensors, comprising  $N_\phi,  N_\theta$ sensors in the toroidal $\phi$ and poloidal $\theta$ directions, respectively. Let us flatten the predicted array of signals $F(\theta, \phi) = \bf{B} \cdot \bf{\tau}$ into a 1D array $\mathbf{F}=\{F_1, ..., F_k, ...., F_N\}$ where each element has a unique $\theta$ and $\phi$: $F_k= F(\theta_i, \phi_j)$. Here, we have chosen the ordering  $k=i + (j-1) N_\phi$. We are given an array of signals $\mathbf{X}=\{X_1 .. X_N\}$, each of which corresponds to the sum of an unknown element of F with an  an arbitrary sign and noise $x_{noise}$. That is, 
\begin{equation}
X_i = \sigma_i \left ( { F_{\pi(i)} + \eta n_i } \right ), \label{eq:Xsig}
\end{equation}
where $\sigma_i \in \{ -1,1\} $, $\pi(i)$ is an unknown permutation of the set $\{1, ... , N\}$, $\eta$ the signal-to-noise ratio, and $n_i$ the signal noise.  

\subsection{Single active coil fit}
The goal is to determine a sorted index list $i_{sort}$ and vector of signs $s_k = \pm 1$, such that we minimise the total mean squared error: 
\begin{equation}
\varepsilon=\sum_{k=1}^N \left ( {s_k X_k - F_{i_{sort}(k)}} \right)^2.    \label{eq:min_single}
\end{equation}
Here, the index $i_{sort}(k)$ is required to be unique and exhaustive.  Our goal is to determine the vectors $I_{sort}$ and $S_{sort}$ of indices $i_{sort}$ signs $s_k$ that minimise Eq. (\ref{eq:min_single}) and returns the corrected signal 
\begin{equation}
Y(i_{sort}(k)) = s_k X_k.  \label{eq:Y_soln}
\end{equation}
The vector $I_{sort}$ is a bijection,
\begin{equation}
    I_{sort}:\{1,...,N\} \rightarrow \{1,...,N\}.
\end{equation}
Formally, the problem can be written as solving for $(I_{sort}, S_{sort})$:
\begin{equation}
(I_{sort}, S_{sort}) =\arg \min_{I,S} \sum_{k=1}^N \left ( {s_k X_k - F_{i_{sort}(k)}} \right)^2, \label{eq:min_single_sort}
\end{equation}
such that $(I_{sort}, S_{sort})=\arg \min_{I,S} \varepsilon(I,S)$ (\textit{ie.} the choice of permutation and signs that minimises the error). The optimisation assumes independent additive noise and uniform weighting across sensors.  A successful reconstruction is defined as exact recovery of both the permutation vector and sensor polarity vector.

There are $N!$ possible coil positions and $2^N$ possible polarities. A Brute force permutation of all combinations (computing $\varepsilon$ for all combinations) scales as $O(N! 2^N)$, which rapidly becomes infeasible. For $N=10$, $N! 2^N = 3.7 \times 10^9 $, which is already extremely computationally demanding.  An alternate optimisation / identification strategy is desirable. 

The mathematically formulated optimisation problem was analysed using AI-assisted coding tools, which identified the structure as a signed assignment problem and suggested a Hungarian-algorithm-based optimisation strategy.  AI-assisted code development was used to implement the MATLAB optimisation workflow using \lstinline|matchpairs|, after which the implementation was debugged, validated, and rigorously tested.


The Hungarian algorithm is a systematic method for solving the assignment problem, where you want to match $N$ items in one set to $N$ items in another while minimizing total cost. It works by transforming the original cost matrix into a form where the optimal assignment becomes easy to identify. First, it performs row and column reductions, subtracting the smallest value in each row and column so that every row and column contains at least one zero. These zeros represent candidate assignments with no additional cost relative to the minimum in their row/column.

The algorithm then tries to select a set of independent zeros (no two in the same row or column) that cover all assignments. To guide this, it repeatedly covers all zeros with the minimum number of lines (rows or columns). If the number of lines equals $N$, an optimal assignment can be constructed directly from the zeros. If not, the matrix is adjusted: the smallest uncovered value is subtracted from all uncovered elements and added to elements covered twice. This creates new zeros while preserving the relative structure of the problem. By iterating this process, the algorithm guarantees convergence to the globally optimal assignment in polynomial time, typically $O(N^3)$, making it efficient for moderately sized problems.


As an illustration, we have constructed a test predicted signal $F_k = \cos(0.45 \pi  k/N)$ with $N=15$. We have then constructed the test measure signal $X_k$ using Eq. (\ref{eq:Xsig}) where noise $n_k$ is described by the independent additive superposition:
\begin{equation}
    n_k = n_{th} + n_{amp} + n_{pink} + n_{emi}  \label{eq:noise}
\end{equation}
where $n_{th}=V_{coil} \mathcal{N}(0,1)$ is the thermal noise of a test coil with $V_{coil} = \sqrt{2 k_B T_{coil} R_{coil} F_s}$, 
$n_{amp}= V_{amp} \sqrt{F_s} \mathcal{N}(0,1)$ is the amplifier noise, 
$n_{emi}= V_{emi} \sin(2 \pi 50 f/F_s)$ is the electromagnetic interference hum,
and
$n_{pink}$ is a pink noise representing electronics $1/f$ sampling and simulated as follows: 
\begin{eqnarray}
    \chi_{pink}   &  = & \mathbb{R} 
    \left [ { 
    \mathcal{F}^{-1} \left \{ { 
    \mathcal{F} \{ \mathcal{N}(0,1) \} (1 + f)^{-1/2}
     } \right \} } 
    \right ] \\
    \eta_{pink}   & = & V_{pink} \times \chi_{pink}/\sigma(\chi_{pink})
\end{eqnarray}
Here, $\mathcal{N}(0,1)$ denotes the normal distribution with mean zero and standard deviation 1, $\mathcal{F}$ is the discrete fast Fourier transform, $\sigma$ denotes the standard deviation, and $k_B$ is Boltzman's constant.   
Table \ref{tab:noise} introduces the constants in these definitions and lists the values used for simulation.   The least-squares formulation implicitly assumes independent additive measurement noise and uniform weighting across sensors. 

\begin{table}[h]
\caption{Coil noise parameters, description, and value.}
\centering
\begin{tabular}{l l l}
\hline
Parameter & Description & Value \\
\hline
$ T_{coil} $  & coil temperature        & 300~K \\
$ R_{coil} $  & coil resistance         & 100 $\Omega$\\
$ F_{s} $     & sampling frequency      & 100~kHz \\
$ V_{amp} $   & amplifier noise density & $5 \times 10^{-9}$ V/$\sqrt{\mathrm{Hz}}$ \\ 
$ V_{emi} $   & emi amplitude           & $5 \times 10^{-6}$ V\\
$ V_{pink} $ & emi amplitude            & $1 \times 10^{-6}$ V\\
\hline
\end{tabular}
\label{tab:noise}
\end{table}

\begin{figure}
\centering
       \includegraphics[width=0.4\textwidth]{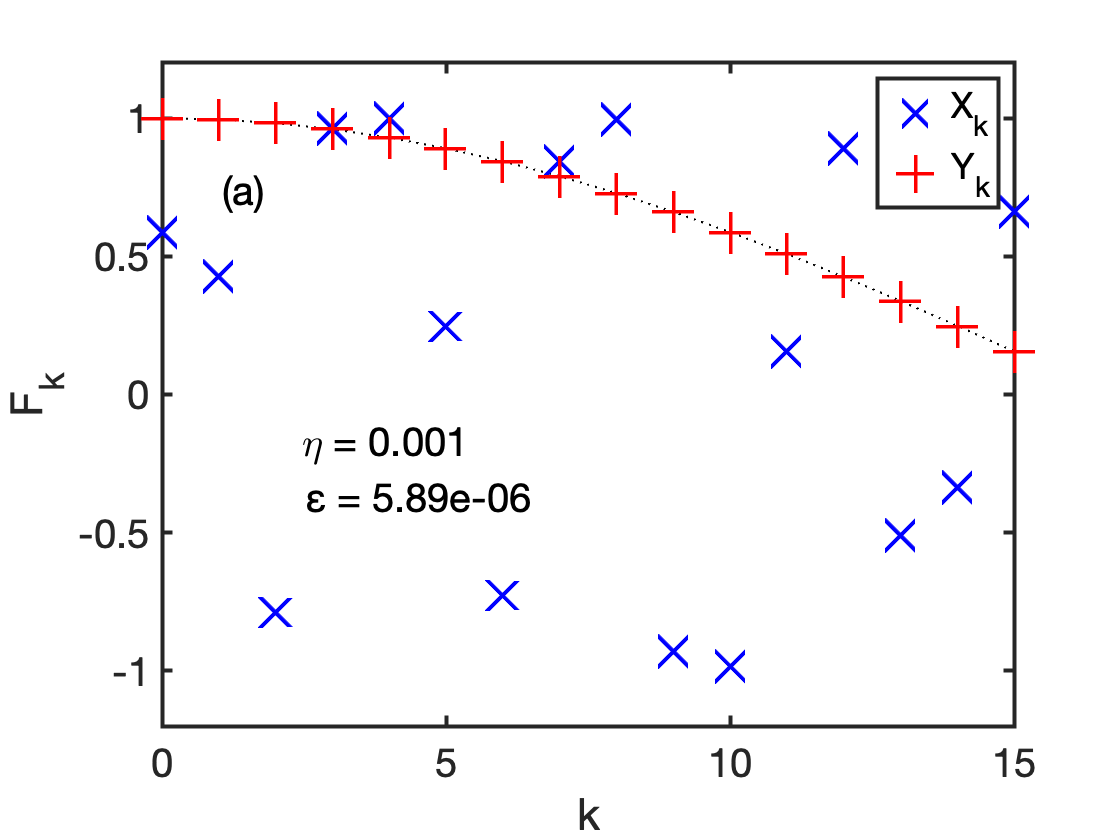}
       \includegraphics[width=0.4\textwidth]{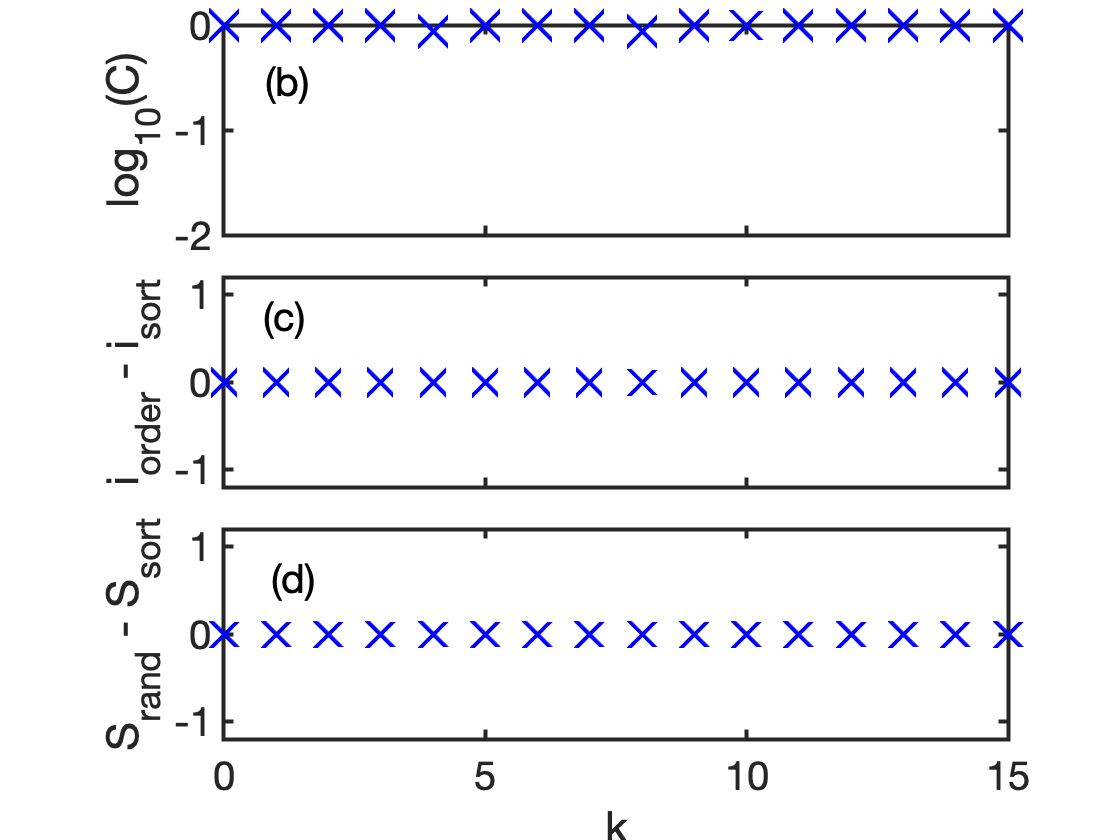}
 \caption{ (a) Test $F_k = \cos(0.45 \pi  k/N)$ and one realisation of the test measured signal $X$, with $\eta = 0.001$.
 (b) confidence metric $C(k) = 1-\varepsilon_k^{(1)}/\varepsilon_k^{(2)}$ representing a measure of confidence that the match is robust ($C(k)=1$ means the match is robust)  
 (c) the index difference  $\pi(k)- i_{sort}(k)$ (zero for the correct fit)
 (d) the polarity difference $s_k-\sigma_k$ (zero for the correct fit) }
\label{fig:test1_vareps0.001}
\end{figure}

Figure \ref{fig:test1_vareps0.001} presents the tested predicted signal $F_k = \cos(0.45 \pi  k/N)$ and one realisation of the test measured signal $X$, with $\eta = 10^{-3}$.  For this test the total error is $\varepsilon \approx 6 \times 10^{-6}$.  Also shown is the sorted and polarity corrected signal $Y_k$, and in panels (b)-(d) the difference  $\pi(k)- i_{sort}(k)$ and $s_k-\sigma_k$ computed by the search algorithm, and the confidence parameter  
\begin{equation}
    C(k) = 1-\varepsilon_k^{(1)}/\varepsilon_k^{(2)}
\end{equation}
where $\varepsilon_k^{(1)}$ is the best (least) assignment cost for optimal
assignments $i_{sort}^{(1)}(k)$ and $S_{sort}^{(1)}(k)$, and   $\varepsilon_k^{(2)}$ the next-best (least) assignment cost for assignments $i_{sort}^{(2)}(k)$ and $S_{sort}^{(2)}(k)$.  The confidence parameter therefore measures the separation between the optimal and next-best admissible assignments.  If $C(k) \approx 1$ the match is robust, or if $C(k)\ll 1$ the match is near-degenerate, and easily swappable with some other pair.  For this case the algorithm returns correct identification of all coil indices and polarity, with $C>0.88$ and global error $\varepsilon \approx 6 \times 10^{-6}$.  The matlab cpu execution time for complete sorting is 0.06~s. 

\begin{figure}
 \centering
       \includegraphics[width=0.4\textwidth]{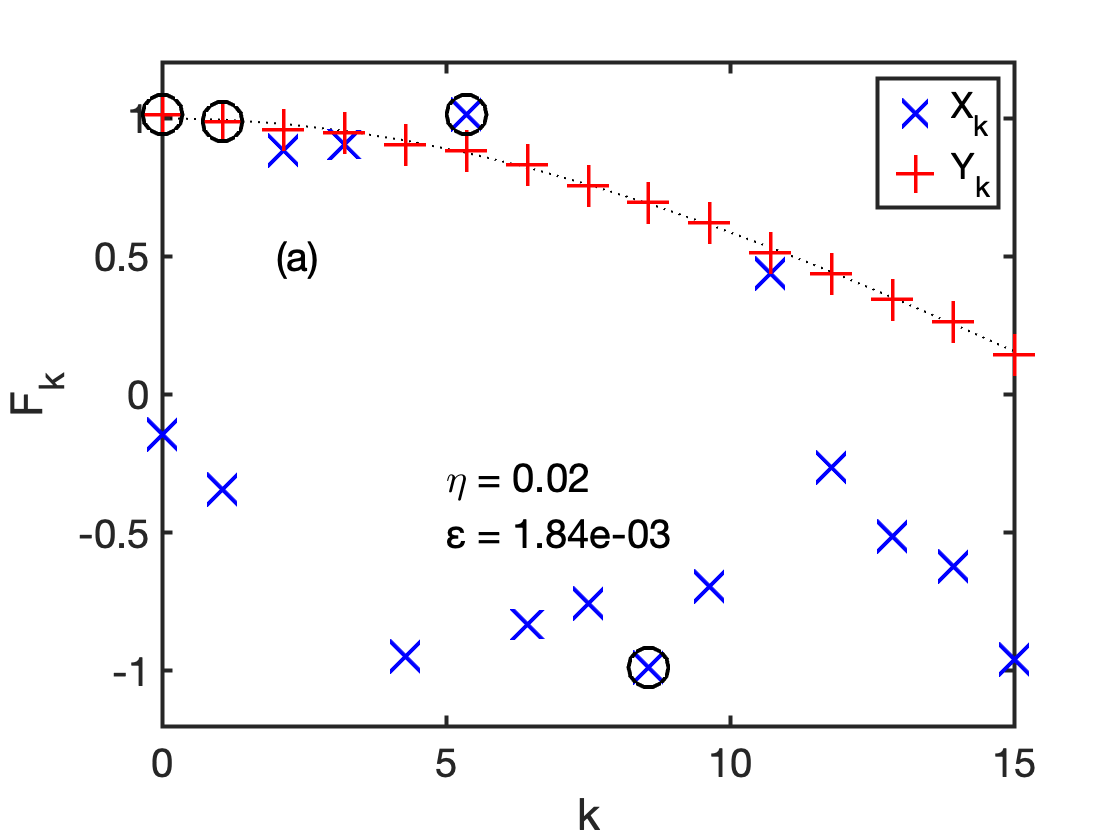}
       \includegraphics[width=0.4\textwidth]{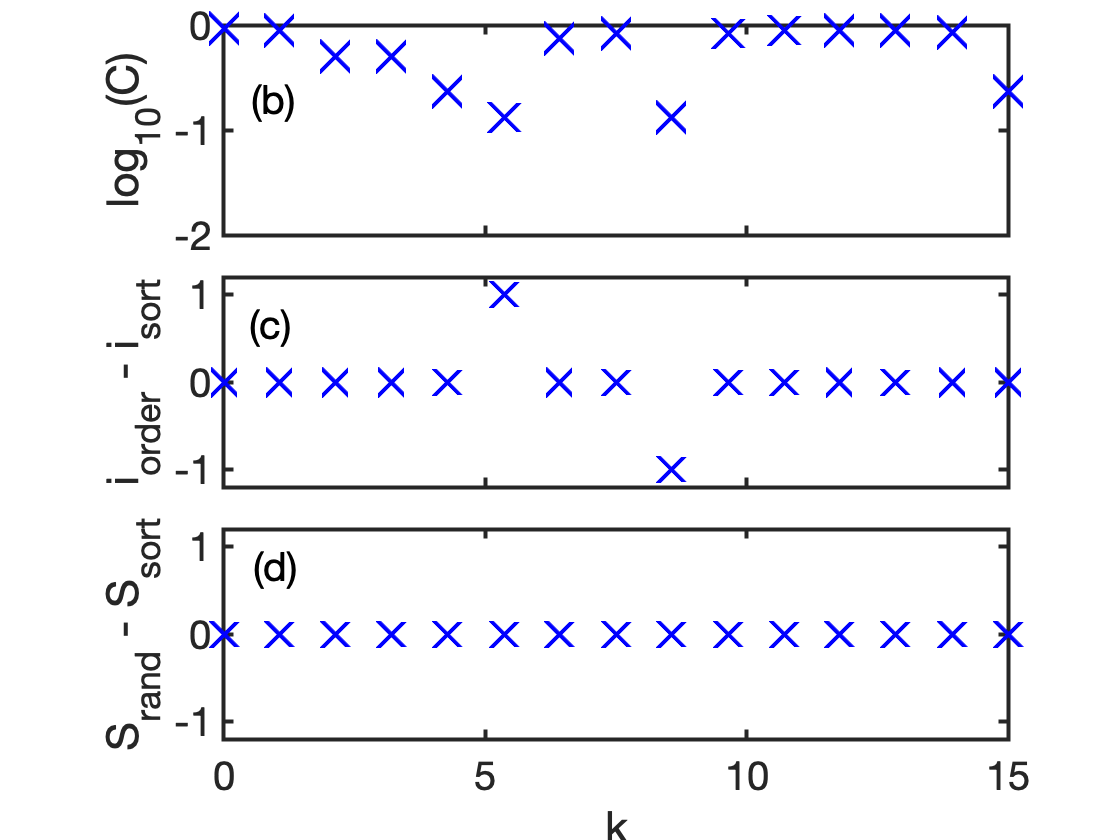}
 \caption{ (a) Test $F_k = \cos(0.45 \pi  k/N)$ and one realisation of the the test measured signal $X$, with $\eta = 0.02$.
 (b) confidence metric $C(k) = 1-\varepsilon_k^{(1)}/\varepsilon_k^{(2)}$ 
 (c) the index difference  $\pi(k)- i_{sort}(k)$
 (d) the polarity difference $s_k-\sigma_k$. 
 In panel (a) solutions which have failed the matching algorithm are circled. } 
\label{fig:test1_vareps0.2}
\end{figure}

Figure \ref{fig:test1_vareps0.2} shows the results for $\eta = 0.02$, with significantly worse total error $\varepsilon 1.3 \times 10^{-3} $. In this case the algorithm has failed, swapping indices at $k=6$ and $k=9$ where at $C(k) \approx 0.1$. These indices correspond to $\pi(6)=2, \pi(9)=1$, and have been circled in panel (a).  The execution time is 0.02~s.

\begin{figure}


 \centering
       \includegraphics[width=0.4\textwidth]{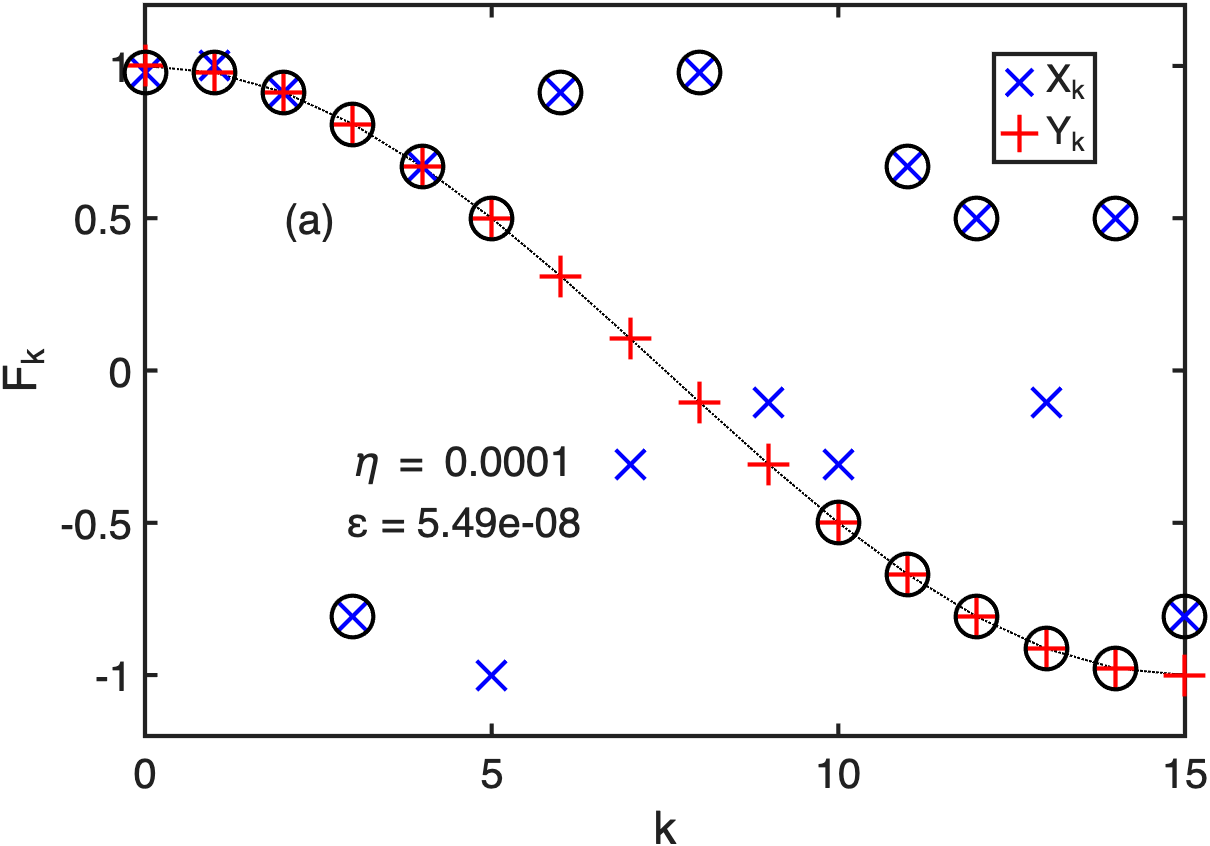}
       \includegraphics[width=0.4\textwidth]{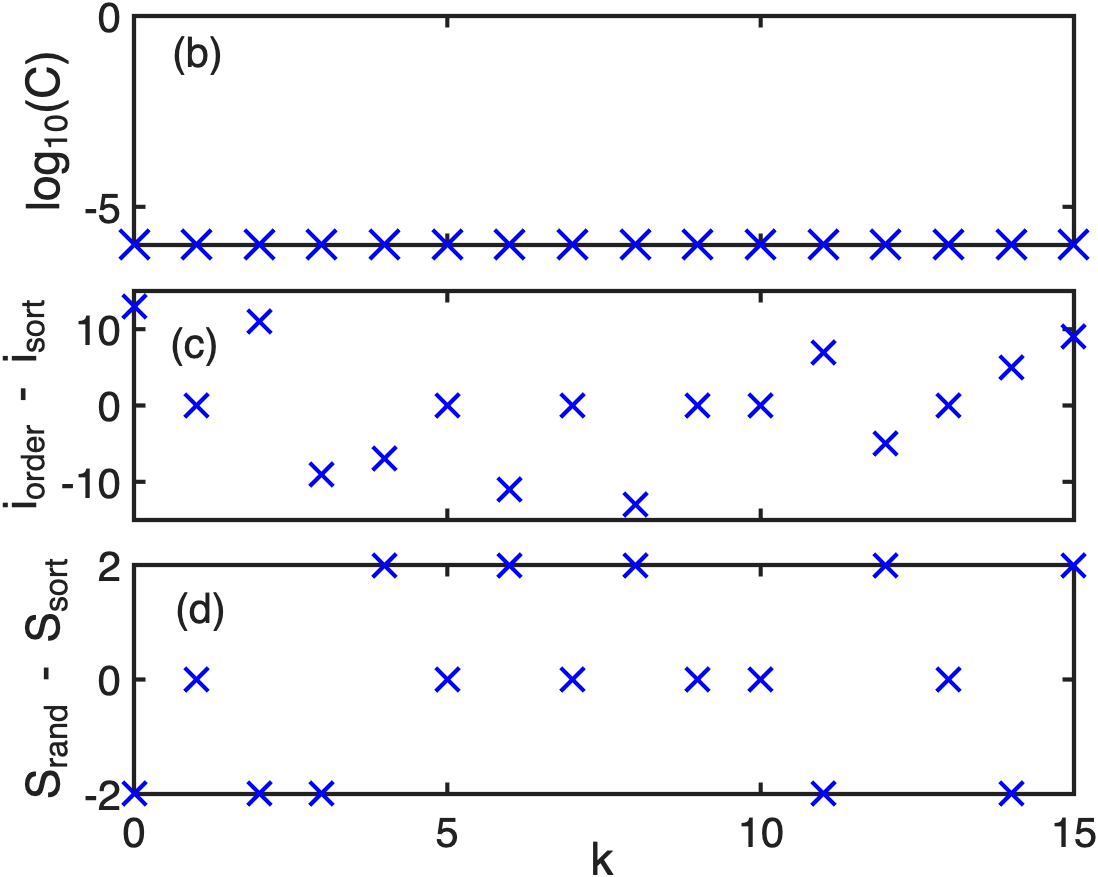}
 \caption{ (a) Test $F_k = \cos( \pi  k/N)$ and one realisation of the the test measured signal $X$, with $\eta = 0.0001$.
 (b) confidence metric $C(k) = 1-\varepsilon_k^{(1)}/\varepsilon_k^{(2)}$ 
 (c) the index difference  $\pi(k)- i_{sort}(k)$
 (d) the polarity difference $s_k-\sigma_k$. 
 In panel (a) solutions which have failed the sorting algorithm are circled. } 
\label{fig:test2_vareps0.0001}
\end{figure}

While successfully demonstrated for $F_k>0$, degeneracy can be illustrated by simple extension across the interval $-1 < F_k<1$. Figure \ref{fig:test2_vareps0.0001} shows test predicted signal $F_k = \cos(\pi  k/N)$ with $N=15$ and one realisation of the test measured signal $X$, with $\eta = 10^{-4}$.  For this test the total error is $\varepsilon = 5.2 \times 10^{-8}$. Again, we have identified solutions for which the algorithm has failed, such that either $i_{order} \ne i_{sort}$ or  $S_{rand} \ne S_{sort}$. However, for all cases for which the algorithm has succeeded, $C(k) < 10^{-6}$, which we have set to $C(k) = 10^{-6}$. This demonstrates the match is not only not robust, it is degenerate. That is, there are equally good alternative fits.  This is not surprising: if both $\pm |F_k|$ are allowed then the algorithm will be unable to disambiguate the sign $\sigma_k$.  

The result shows that it is the confidence metric $C$ that is a much better indicator of robustness (and correctness) of the assignment, not the fit quality as measured by the total error $\varepsilon$. 

\subsection{Multiple active coil fit}

The single coil disambiguation algorithm works providing that the elements of $F_k$ are not degenerate and the signal-to-noise ratio large. The actual field realizations may contain degenerate or near-degenerate signal values. There are however multiple different field  solutions for $F_k$ corresponding to different active field coils, and one can measure the field $X_k$ for each field coil solution.  The sorted vector list $i_{sort}$ and $S_{sort}$ must also work for all $F_k$ and $S_k$. 


The solution is instead to find $(I_{sort},S_{sort})$ to minimise the combined cost 
\begin{equation}
\varepsilon=\sum_{t}^{N_t} \sum_{k=1}^{N_k} \left ( {s_k X_{k,t} - F_{i_{sort}(k),t}} \right)^2,    \label{eq:min_multiple_vareps}
\end{equation}
where there are now $N_t$ active coils. While individual field realizations may admit degenerate signed permutations, the intersection of multiple independent field solutions strongly constrains the admissible assignment space.  Formally, the problem can be written as:
\begin{equation}
(I_{sort}, S_{sort})=\arg \min_{I,S} \sum_{t}^{N_t} \sum_{k=1}^{N_k} \left ( {s_k X_{k,t} - F_{i_{sort}(k),t}} \right)^2.    \label{eq:min_multiple_sort}
\end{equation}
Modification to the algorithm is straight-forward, and amounts to replacement of the cost function in \lstinline|mathpairs|. 

\begin{figure}


 \centering
       \includegraphics[width=0.4\textwidth]{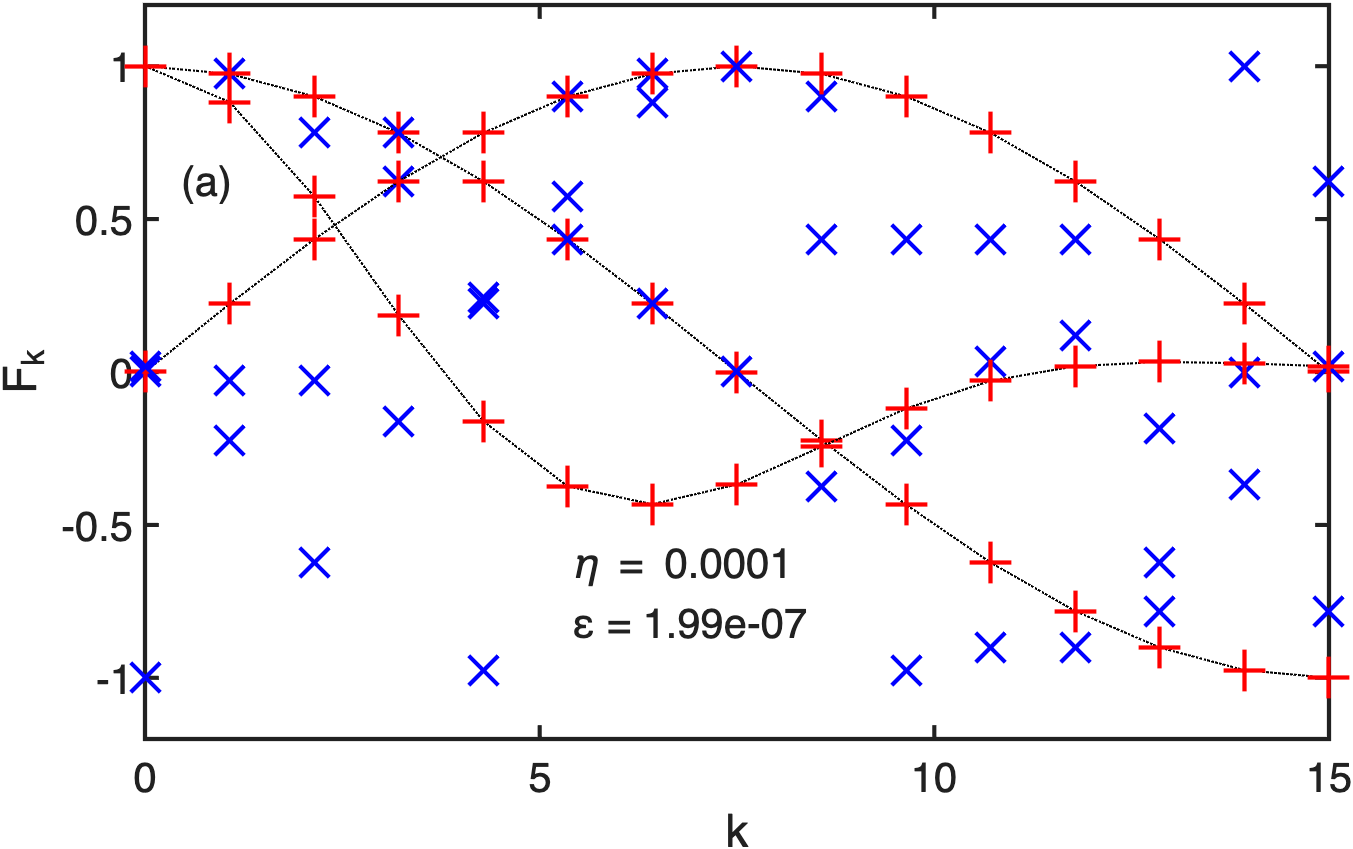}
       \includegraphics[width=0.4\textwidth]{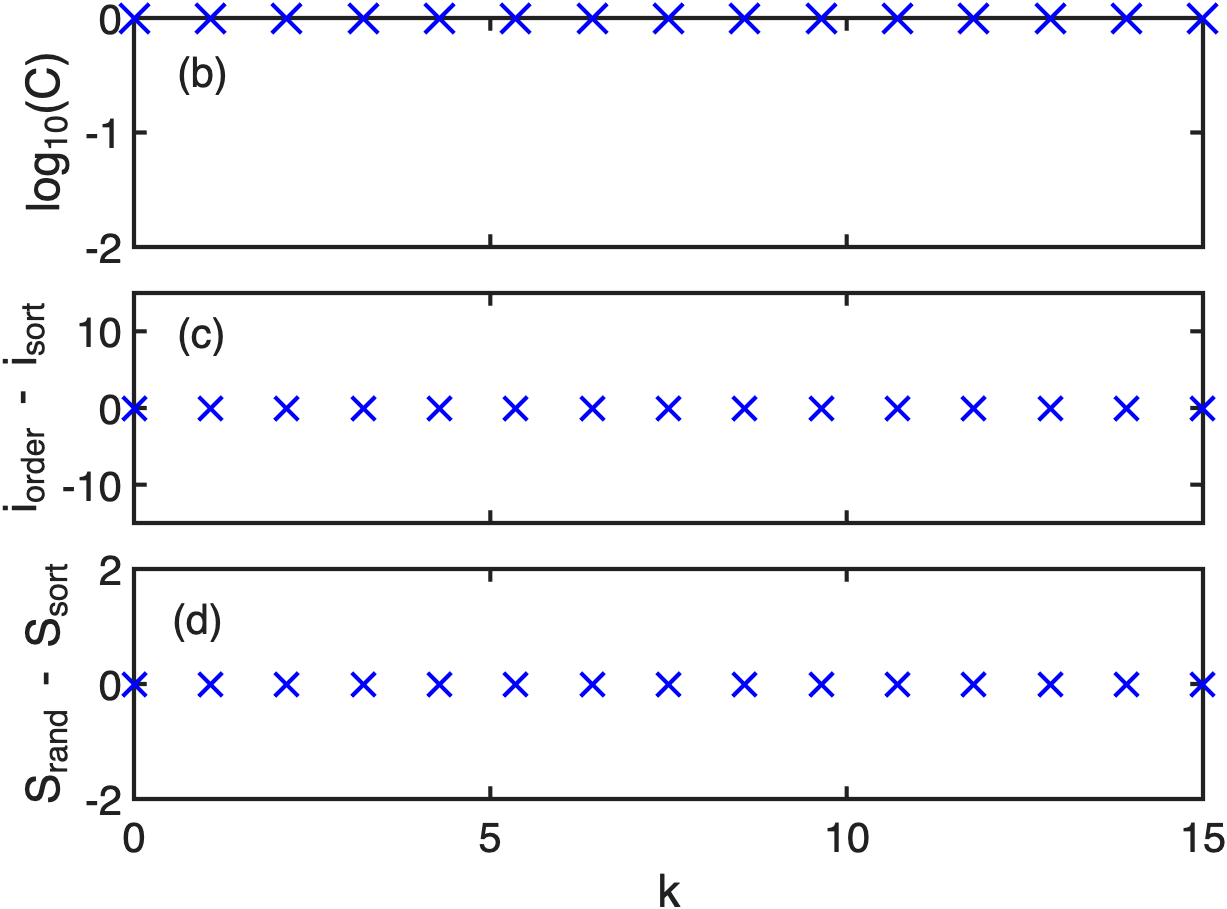}
 \caption{ (a) Test $F_k$ given by Eqs. (\ref{eq:Fk1_cos})-(\ref{eq:Fk3_cosexp}) and one realisation of the the test measured signal $X$, with $\eta = 0.0001$.
 (b) confidence metric $C(k) = 1-\varepsilon_k^{(1)}/\varepsilon_k^{(2)}$ 
 (c) the index difference  $\pi(k)- i_{sort}(k)$
 (d) the polarity difference $s_k-\sigma_k$. 
 } 
\label{fig:test3_vareps0.0001}
\end{figure}

As an illustration, we have constructed test predicted signals:
\begin{eqnarray}
    F_{k,1} & = & \cos(\pi  k/N)  \label{eq:Fk1_cos} \\
    F_{k,2} & = & \sin(\pi  k/N)  \label{eq:Fk2_sin} \\
    F_{k,3} & = & \cos(2 \pi  k/N) \exp{ \left ( { - k^2/(4 N^2)} \right ) } \label{eq:Fk3_cosexp} 
\end{eqnarray}
and repeated the search. Figure \ref{fig:test3_vareps0.0001} shows test predicted signal $F_k$ given by Eqs. (\ref{eq:Fk1_cos} - \ref{eq:Fk3_cosexp} ) with $N=15$ and one realisation of the test measured signal $X$, with $\eta = 10^{-4}$.  For this test the total error is $\varepsilon = 5.2 \times 10^{-8}$.  The algorithm completely disambiguates the signal set, and the global confidence metric is 1, indicating the fit is robust. 

\begin{figure}


 \centering
       \includegraphics[width=0.4\textwidth]{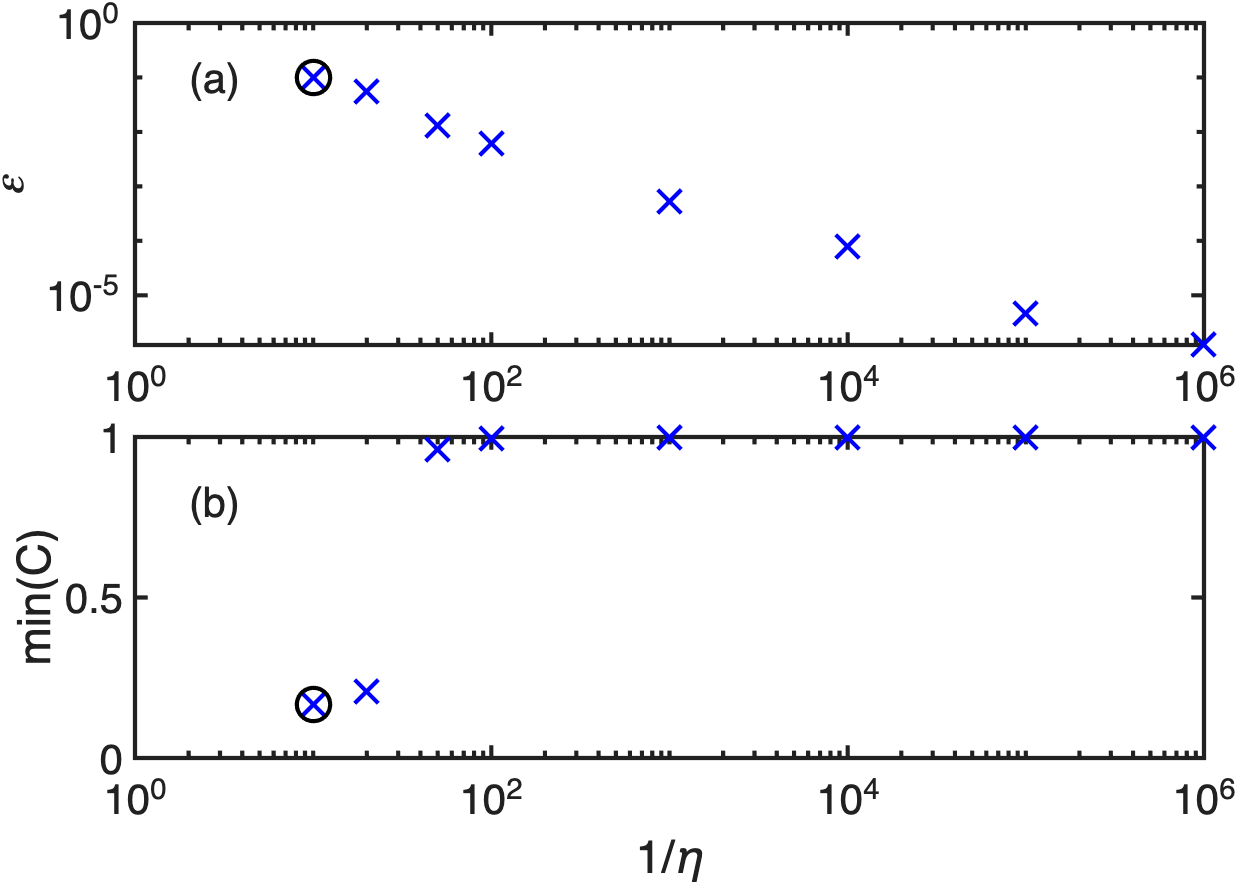}
 \caption{ (a) Residual error $\varepsilon$ and (b) Minimum in confidence as a function of signal-to-noise ratio $1/\eta$.  Here the test function $F_k$ is given by multiple field coils of Eqs. (\ref{eq:Fk1_cos})-(\ref{eq:Fk3_cosexp}). Solutions that have failed the sorting algorithm (have incorrect parity and/or index order) are circled.  } 
\label{fig:test3_scan}
\end{figure}

We have decreased the signal-to-noise ratio (increased $\eta$) and recomputed the fit.  Figure \ref{fig:test3_scan}  shows the variation of the residual error (which measures fit quality) and the minimum global confidence (which measures assignment uniqueness) as the signal-to-noise ratio $1/\eta$ is varied. Results that fail the sorting algorithm have been circled.  The displayed results took less than 40ms per test.  We have applied a time out of 5s on \lstinline|matchpairs|, returning NaN for these cases.  Correct sorting is achieved down to a signal-to-noise ratio of $1/\eta=20$.


%
%
\section{Application to ITER} \label{sec:ITER_app}

In \S \ref{sec:strategy} we saw that incorrect signed permutations that provide acceptable fits for one field realization are generally inconsistent with additional independent field realizations. This motivates a combined least-squares optimisation across multiple independent fields, each possessing high spatial variation and sufficiently large projected field amplitudes across the sensor set. Such an approach rapidly suppresses false assignments and produces solutions with a high confidence of fit. Optimally, one might posit the ideal field realisations have high spatial variation, independent signatures, and few regions where $\mathbf{B} \cdot \tau \approx 0$.  

The results of \S \ref{sec:ITER_coils} show that the different ITER active coil systems generate markedly different spatial structures in the tangential magnetic field component $\mathbf{B} \cdot \tau$ measured by the equilibrium sensors. The axisymmetric PF coils produce vacuum magnetic fields that are toroidally symmetric, while PF1 and PF6 additionally exhibit extended poloidal regions for which $\mathbf{B} \cdot \tau$ is small. The edge control coils produce fields that are localised to the coil, with  $\mathbf{B} \cdot \tau \approx 0$ elsewhere.  In contrast, the correction coils generate strongly non-axisymmetric magnetic structures with high spatial variation in the poloidal and toroidal directions.  In the following section we therefore investigate the use of correction coil field solutions for sensor ordering and polarity identification.  PF coils are added to the combined cost function to improve the confidence of selection for second wall coils. 


We have constructed the 1D vector 
\begin{equation}
F_{k+(l-1) N_l} =F(\theta_k, \phi_l)
\end{equation}
such that the sensors are located at grid points $\theta_1... \theta_{N_k}$ and $\phi_1... \phi_{N_l}$. As illustration we have selected $N_{\theta}=20$ poloidal and $N_{\phi}=20$ toroidal sensors to be regularly spaced across $\{-\pi,\pi\}$ and $\{0,2\pi\}$.  Values of $F(\theta_k, \phi_l)$ were computed by interpolation of a high resolution computation of $\mathbf{B} \cdot \mathbf{\tau}$.  To this, noise was added through Eq. (\ref{eq:Xsig}), and a randomly permuted index $\pi_i$ and a randomly permuted sign $\sigma_i$.  This produces the simulated extraction vector $X_i$.

\begin{figure}


 \centering
       \includegraphics[width=0.4\textwidth]{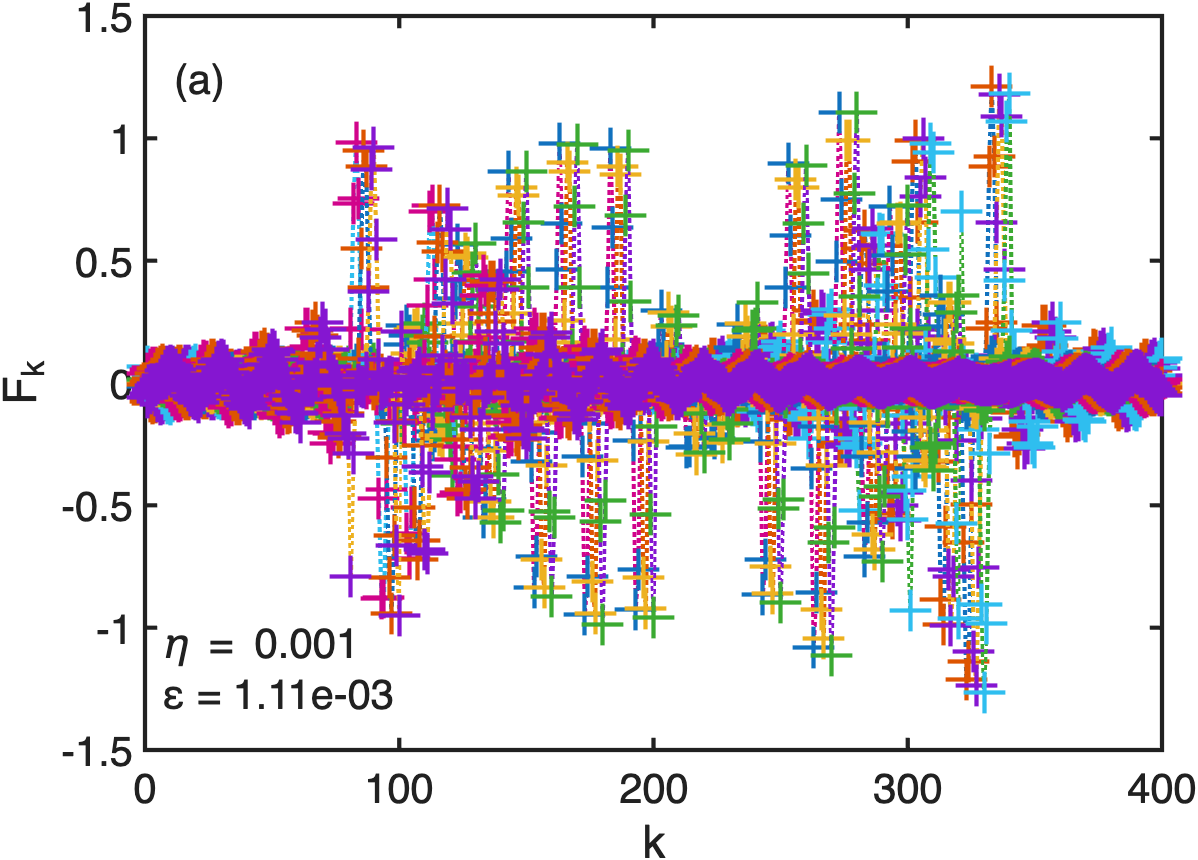}
       \includegraphics[width=0.4\textwidth]{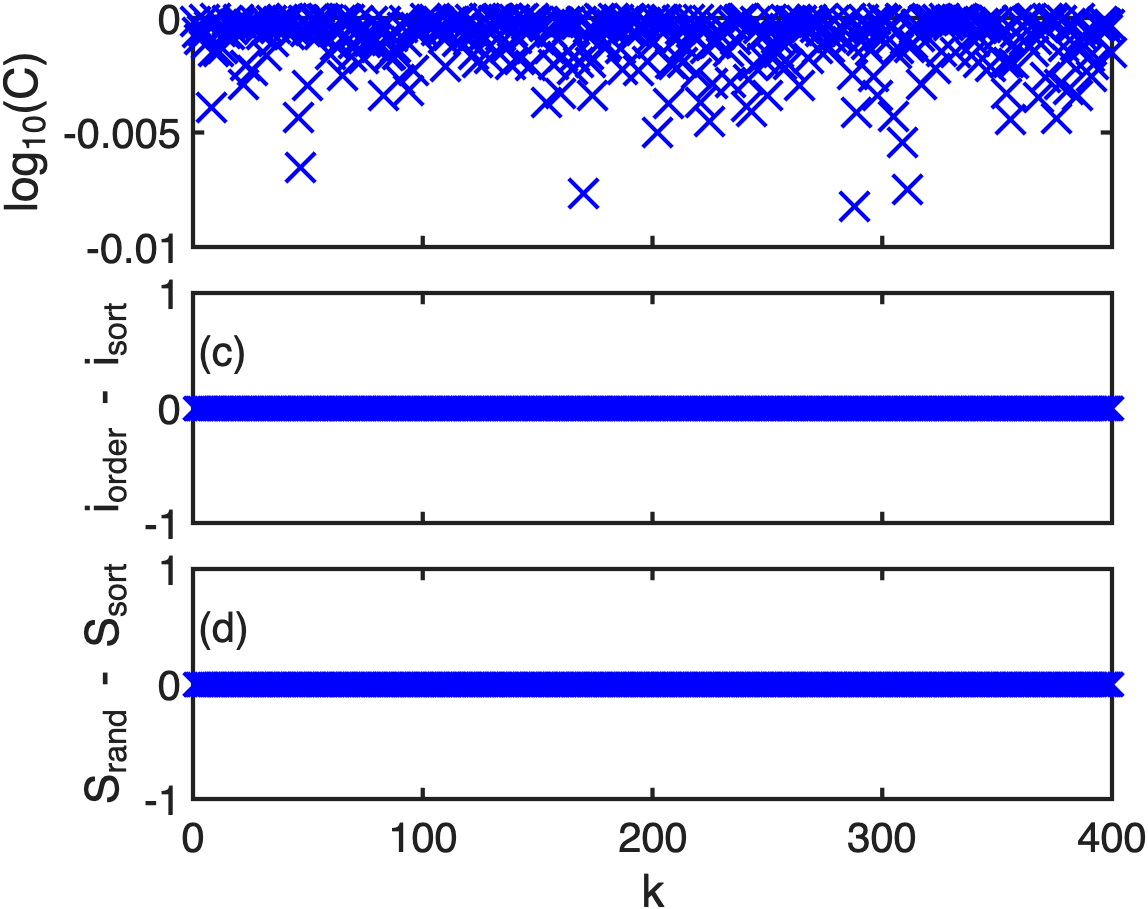}
 \caption{ (a) Test $F_k$ given by $N_{\theta}=N_{\phi}=20$ regularly spaces sensors wrapped into a 1D vector with $\eta = 0.001$.
 (b) confidence metric $C(k) = 1-\varepsilon_k^{(1)}/\varepsilon_k^{(2)}$ 
 (c) the index difference  $\pi(k)- i_{sort}(k)$
 (d) the polarity difference $s_k-\sigma_k$. 
 } 
\label{fig:sample12}
\end{figure}

\begin{figure}


 \centering
       \includegraphics[width=0.4\textwidth]{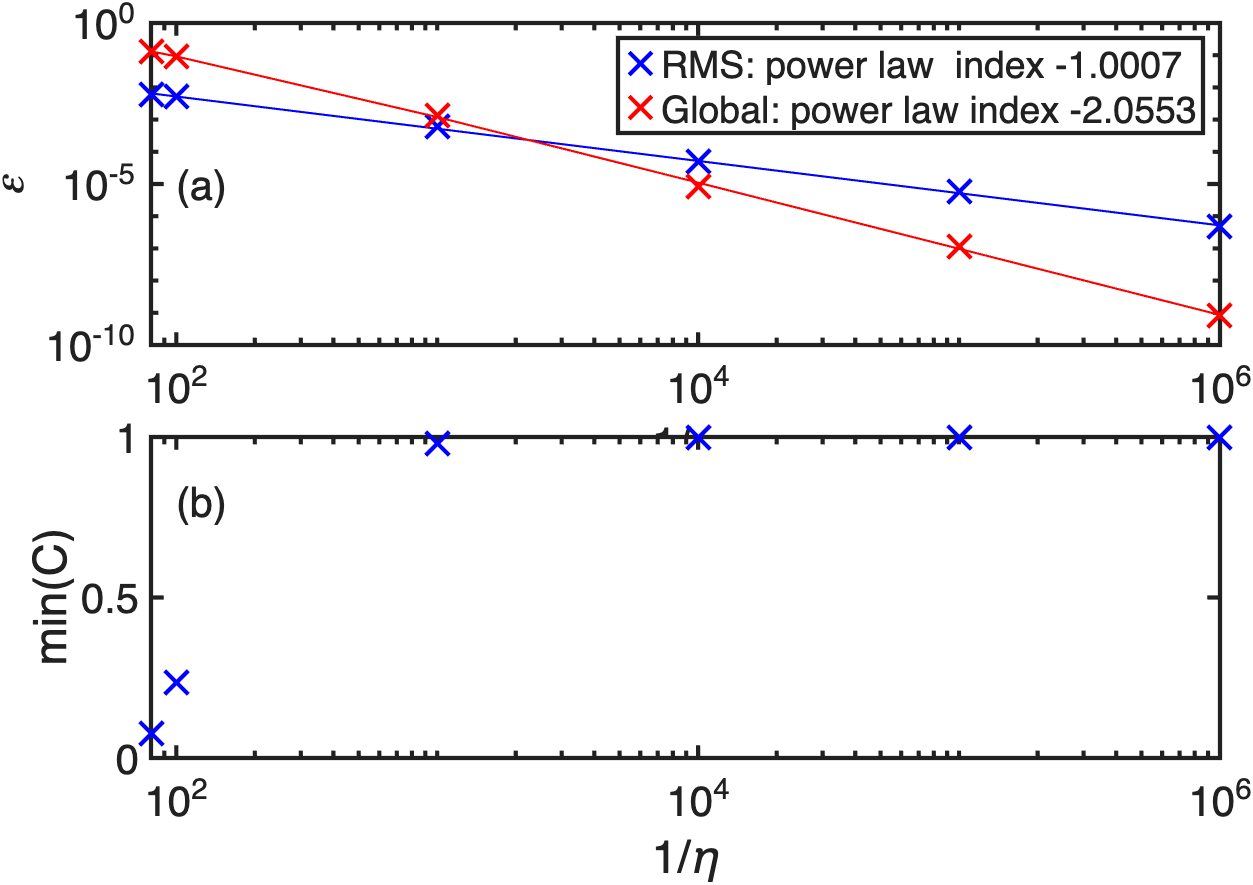}
 \caption{ (a) Residual error $\varepsilon$ and (b) Minimum in confidence as a function of signal-to-noise ratio $1/\eta$.  Solutions that have failed the sorting algorithm (have incorrect parity and/or index order) are circled. 
 } 
\label{fig:sample_fit}
\end{figure}

Figure \ref{fig:sample12} shows the resolved predicted signal $F_k$ for $N_k = N_l=200$ and $\eta=0.001$, together with confidence values of the next fit, and the index and polarity difference. This reveals correct sorting with a confidence above $C\approx 0.99$, suggesting very high measurement uniqueness.  We have varied the signal-to-noise ratio and recomputed the fit. Figure \ref{fig:sample_fit}  shows the variation of the residual error, the global error and the minimum global confidence.  Results that fail the sorting algorithm have been circled.  The displayed results took less than 40ms per test.  We have applied a time out of 5s on \lstinline|matchpairs|, returning NaN for these cases.  Correct sorting is achieved down to a signal-to-noise ratio of $1/\eta=80$. At this value the minimum confidence is less than 0.1.

\begin{figure}
 \centering
       \includegraphics[width=0.4\textwidth]{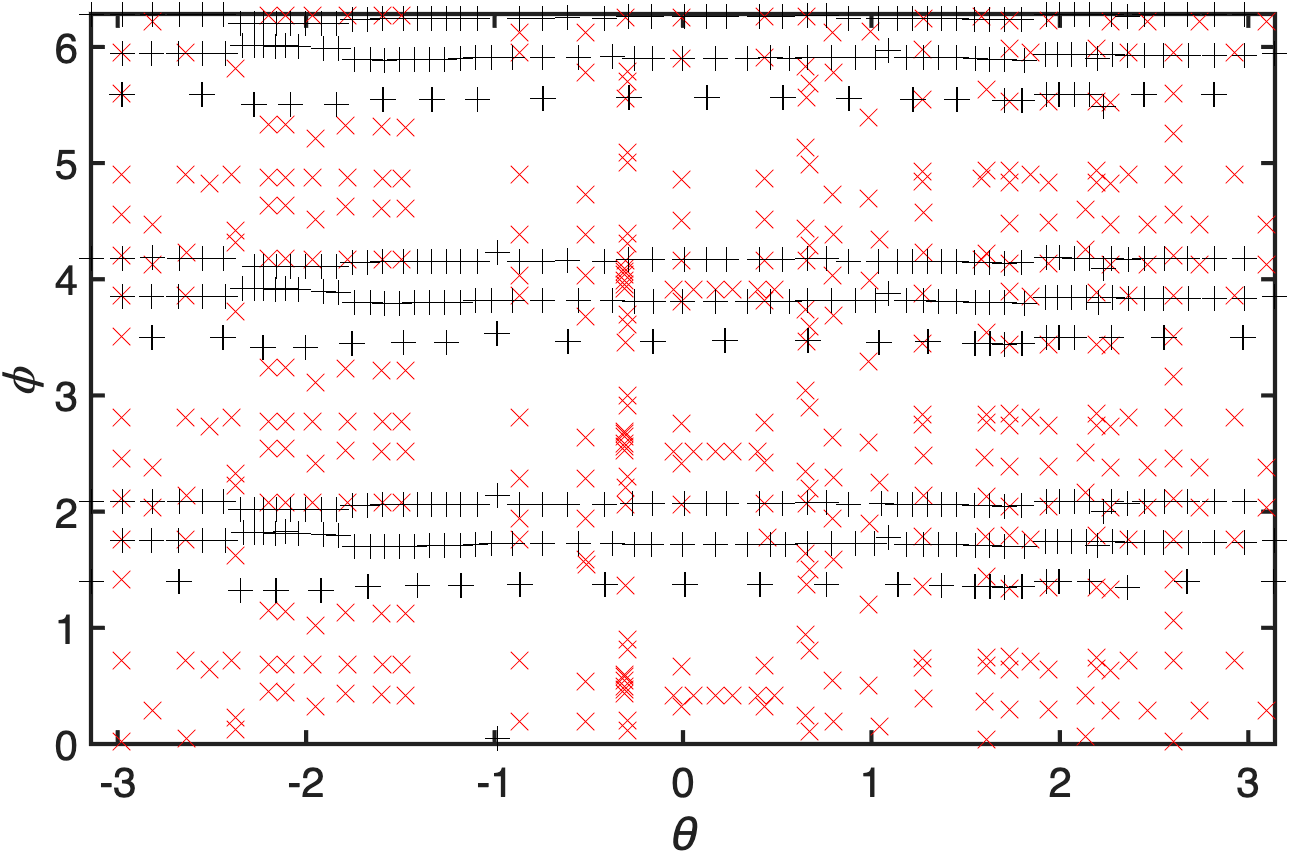}
 \caption{ Projection of poloidal field coils into $(\theta,\phi)$ for first (red) and second (black) wall. 
 } 
\label{fig:bpol_ITER}
\end{figure}

\begin{figure}


 \centering
       \includegraphics[width=0.4\textwidth]{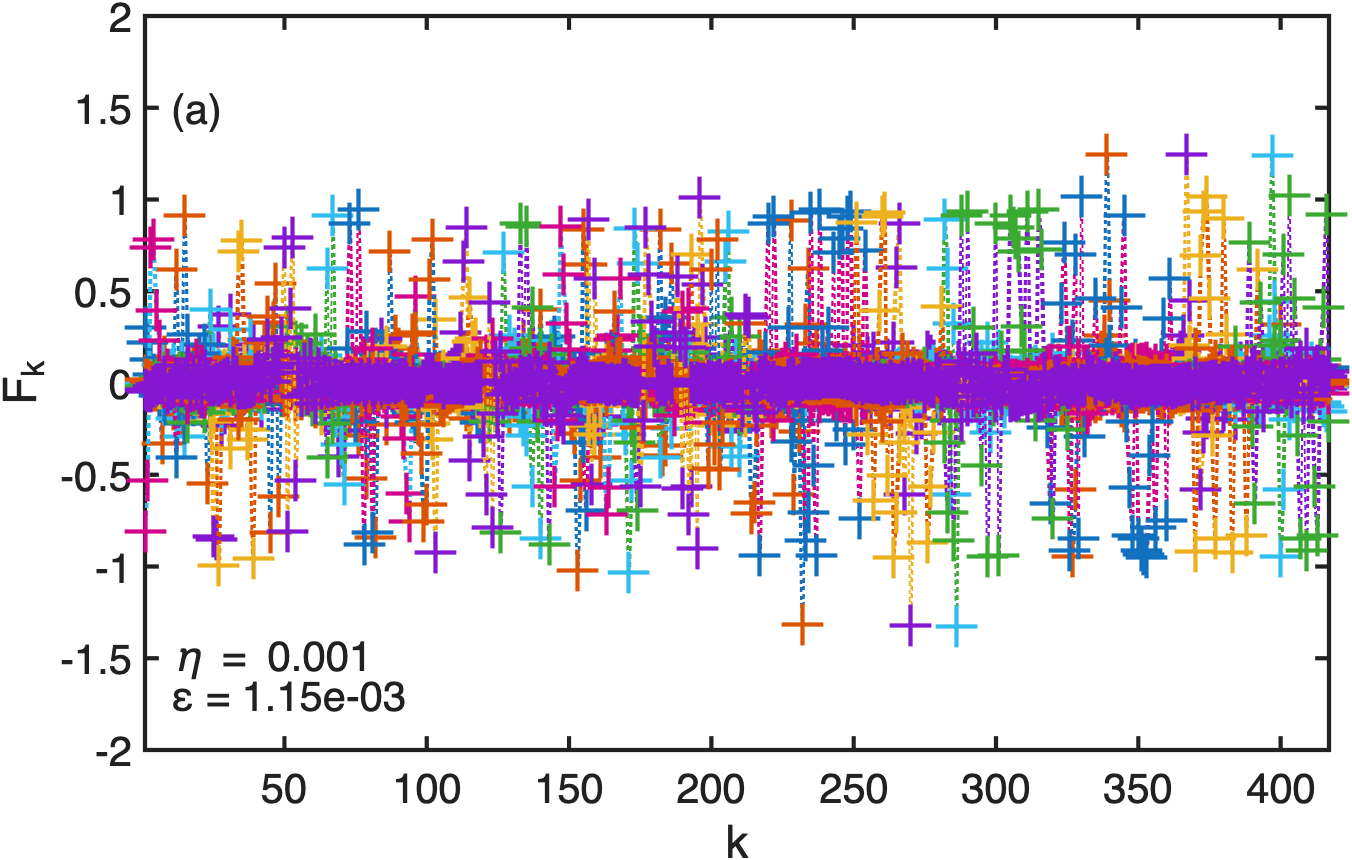}
       \includegraphics[width=0.4\textwidth]{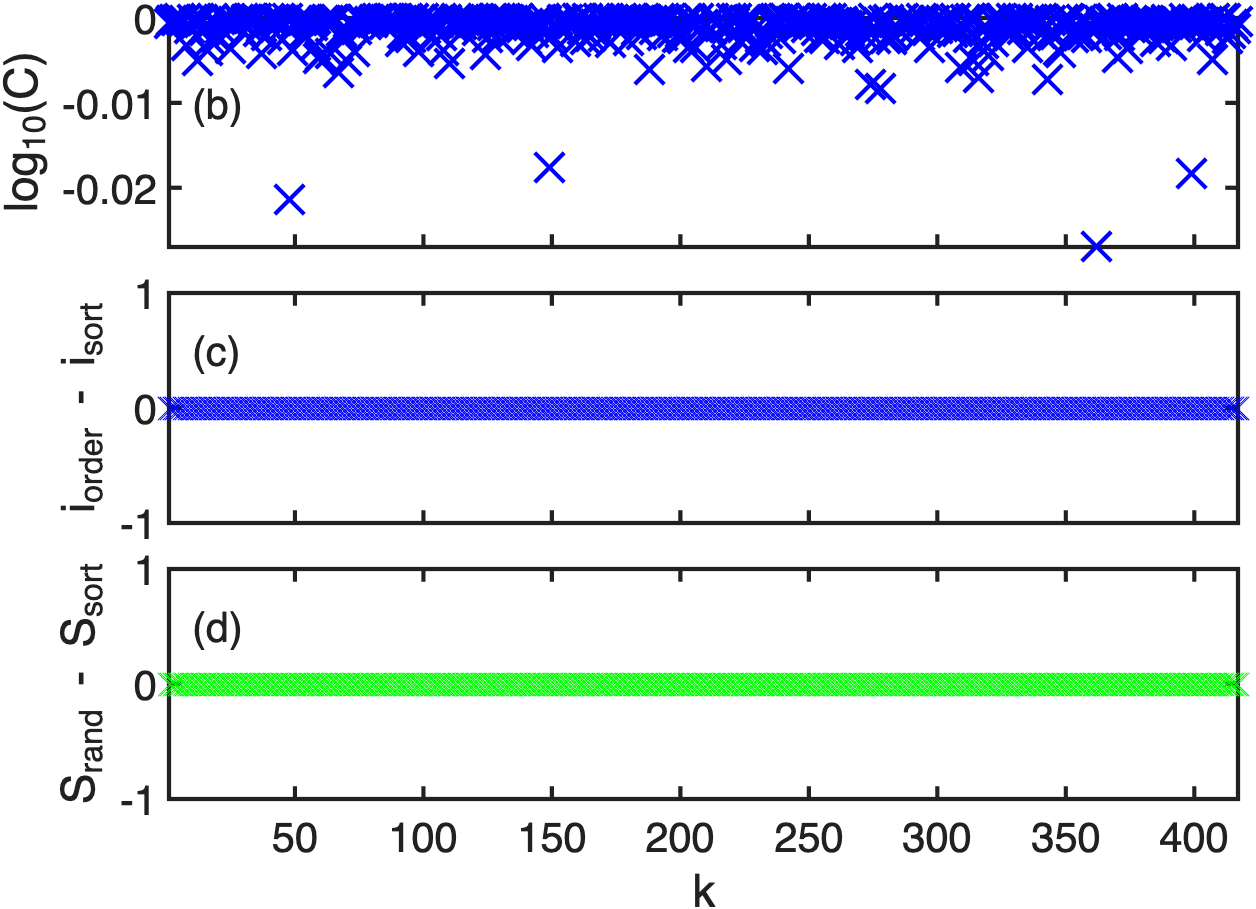}
 \caption{ (a) First wall poloidal field $F_k$ given by sensors wrapped into a 1D vector with $\eta = 0.001$.
 (b) confidence metric $C(k) = 1-\varepsilon_k^{(1)}/\varepsilon_k^{(2)}$ 
 (c) the index difference  $\pi(k)- i_{sort}(k)$
 (d) the polarity difference $s_k-\sigma_k$. 
 } 
\label{fig:ITER_md_wall1}
\end{figure}

\begin{figure}


 \centering
       \includegraphics[width=0.4\textwidth]{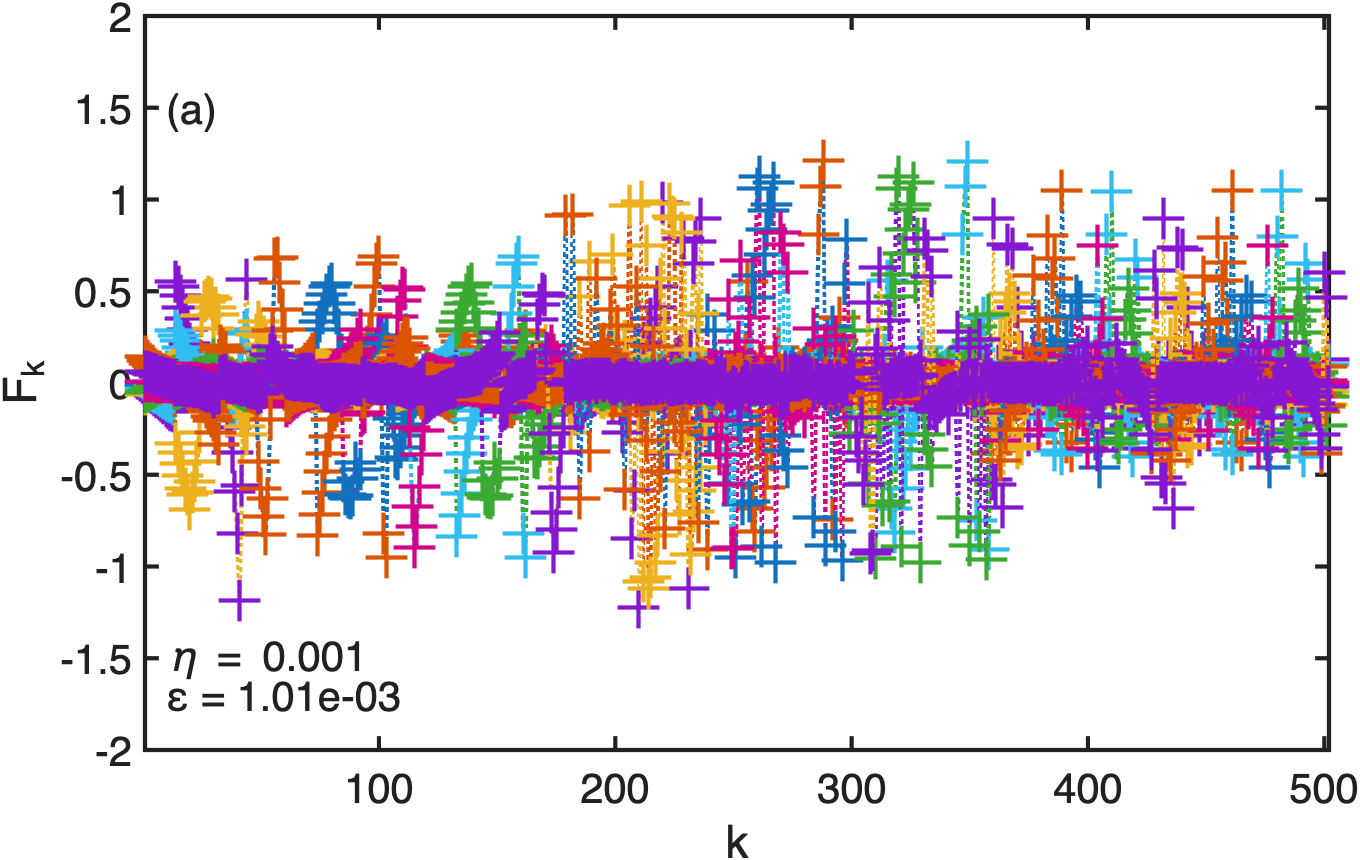}
       \includegraphics[width=0.4\textwidth]{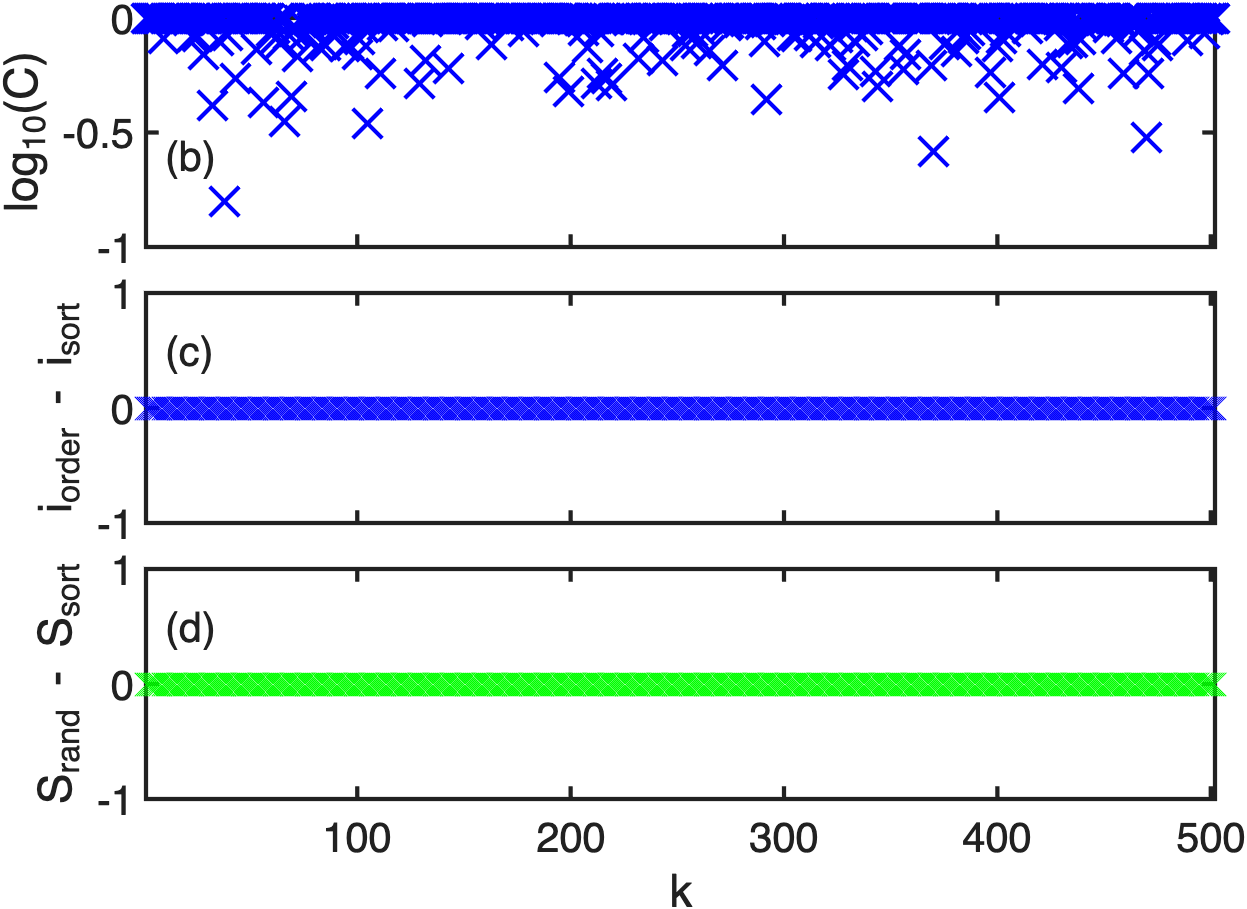}
 \caption{ (a) Second wall poloidal field $F_k$ given by sensors wrapped into a 1D vector with $\eta = 0.001$.
 (b) confidence metric $C(k) = 1-\varepsilon_k^{(1)}/\varepsilon_k^{(2)}$ 
 (c) the index difference  $\pi(k)- i_{sort}(k)$
 (d) the polarity difference $s_k-\sigma_k$. 
 } 
\label{fig:ITER_md_wall2}
\end{figure}

\begin{figure}


 \centering
       \includegraphics[width=0.4\textwidth]{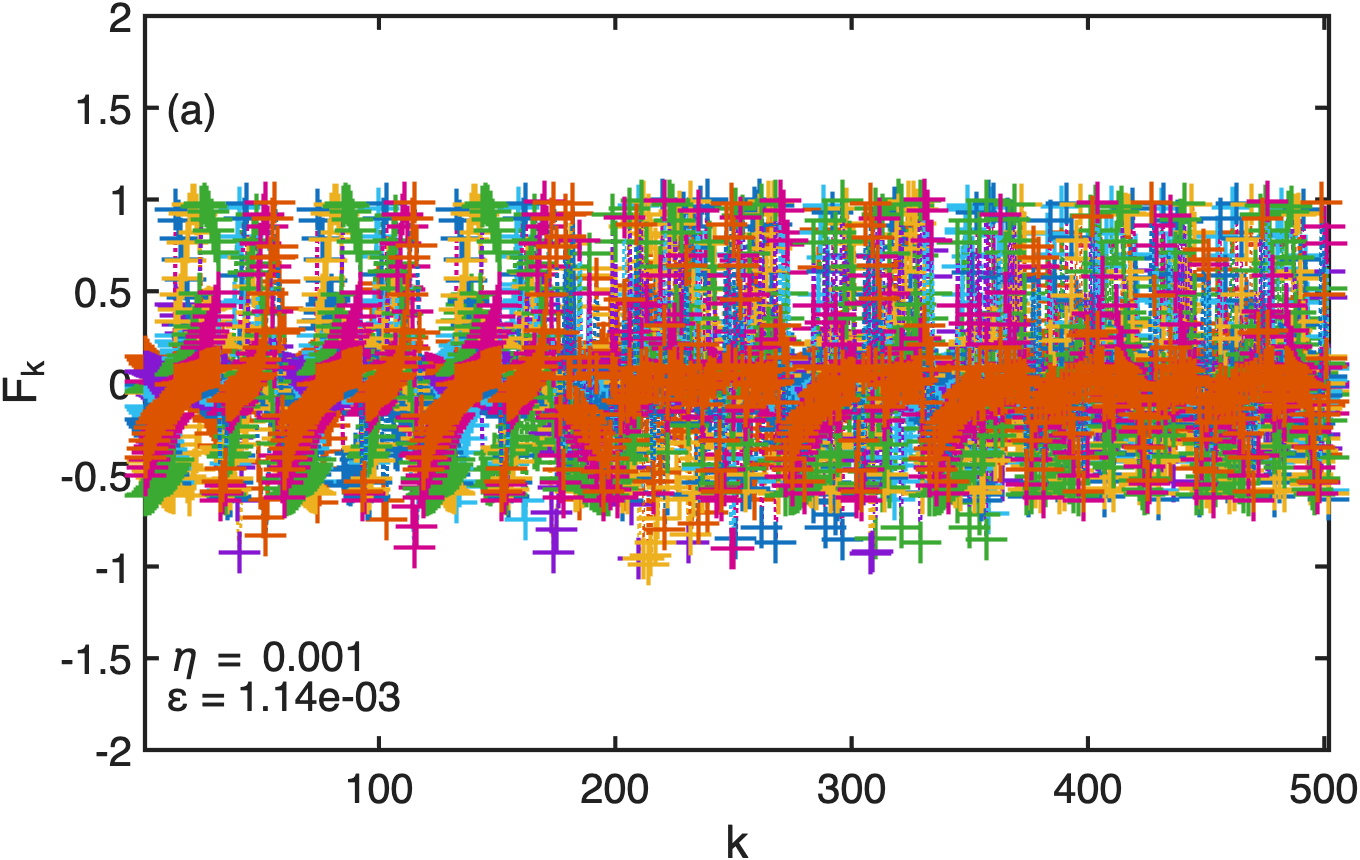}
       \includegraphics[width=0.4\textwidth]{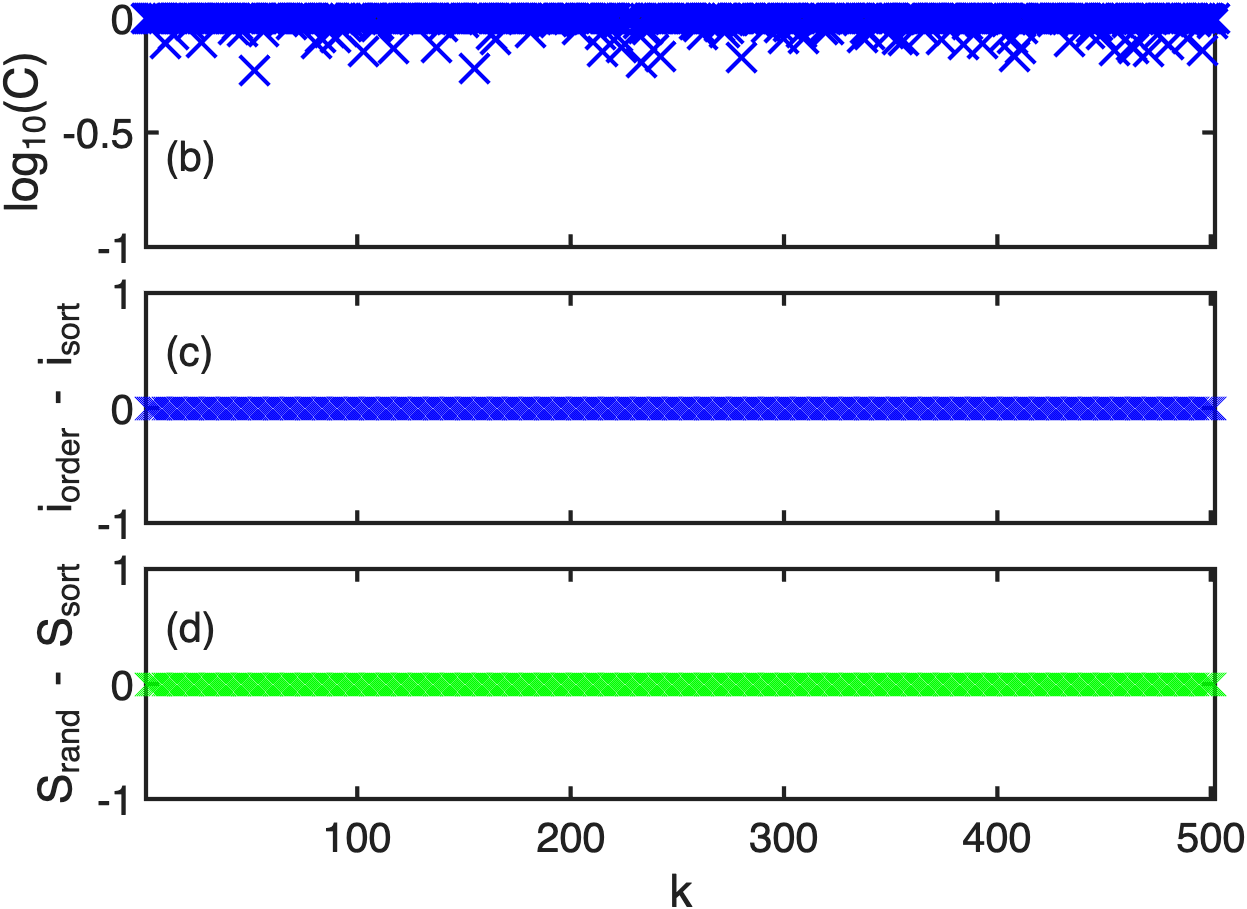}
 \caption{ (a) Second wall poloidal field $F_k$ given by sensors wrapped into a 1D vector with $\eta = 0.001$.
 (b) confidence metric $C(k) = 1-\varepsilon_k^{(1)}/\varepsilon_k^{(2)}$ 
 (c) the index difference  $\pi(k)- i_{sort}(k)$
 (d) the polarity difference $s_k-\sigma_k$. 
 } 
\label{fig:ITER_md_wall2_PF}
\end{figure}

Finally, we have applied the algorithm to the real positions of the poloidal coils in the first and second wall in ITER, as shown in Fig. \ref{fig1:ITER_sensors}(a), and projected into $(\theta, \phi)$ space in Fig. \ref{fig:bpol_ITER}.  Figure \ref{fig:ITER_md_wall1} shows the resolved predicted signal $F_k$ for all 417 poloidal field coils on the first all, and $\eta=0.001$, together with confidence values of the next fit, and the index and polarity difference. This reveals correct sorting with a confidence above $C(k)\approx 0.97$.   The second wall poloidal field coils (see Fig. \ref{fig:bpol_ITER}) comprise 9 arrays at near constant $\phi$, each of which is regularly spaced in $\theta$.  Figure \ref{fig:ITER_md_wall2} shows the resolved predicted signal $F_k$ for all 502 poloidal field coils on the second wall, and $\eta=0.001$, together with confidence values of the next fit, and the index and polarity difference. This reveals correct sorting with a confidence above $C(k)\approx 0.15$. To improve the confidence we have added the active poloidal field coils PF1 - PF6 to the combined cost function of Eq. (\ref{eq:min_multiple_vareps}). The predicted signals, sorted signals and confidence is shown in Fig. \ref{fig:ITER_md_wall2_PF}, with confidence above $C(k)\approx 0.59$. 

The results demonstrate that appropriately chosen active coil excitations can transform the sensor identification problem from a highly degenerate combinatorial search into a well-constrained optimisation problem.

\section{Conclusions} \label{sec:conclusions}

The present work has developed a strategy for determining the polarity and spatial ordering of poloidal magnetic flux coils embedded in the ITER first and second walls. The approach combines energisation of active field coils (aysmmetric correction coils and poloidal field coils), computation of the projected magnetic field at the wall, and formulation of the reconstruction as a signed assignment problem. AI-assisted coding tools identified the assignment structure and suggested a Hungarian-algorithm-based optimisation strategy, after which the implementation was independently debugged, validated, and rigorously tested. A confidence metric was also developed to quantify the robustness and uniqueness of the reconstructed assignments. 

The Hungarian search algorithm scales as $O(N^3)$, yet remained computationally tractable for large sensor sets. In synthetic tests involving 400 regularly spaced poloidal flux coils, robust reconstruction was achieved down to a signal-to-noise ratio of approximately 80, with a total computation time of approximately 2 s. The method was subsequently applied to the first- and second-wall flux coil arrays of ITER.  Using the correction coils alone, the first wall coils were reconstructed with high confidence, $C(k)>0.97$.  

Reconstruction of the second-wall arrays, which consist of multiple sensor sets distributed at nearly constant toroidal angle and regularly spaced in poloidal angle, proved substantially more degenerate, yielding confidence values of only $C(k)>0.15$.  Incorporating the active poloidal field coils into the combined least-squares cost function significantly improved the reconstruction confidence to $C(k)>0.59$.   

The results demonstrate how spatially distinct active magnetic perturbations can be used to systematically break combinatorial degeneracies in large magnetic diagnostic systems.  It would be interesting to study how reconstruction performance degrades as the inverse problem becomes progressively noise dominated.  This could be investigated by a Monte Carlo analysis in which disambiguation is repeated for multiple noise realisations, and the probability of successful sensor disambiguation examined as a function of signal-to-noise ratio.

The assignment formulation naturally enforces uniqueness constraints while remaining computationally tractable for large sensor sets.  Future works could compare this approach with alternative inference frameworks, such as a maximum-correlation \cite{Kay_1993}, probabilistic inference \cite{Bishop_2009} or Bayesian inverse-problem formulations \cite{von_Nessi_2014, Tarantola_2005}. 

The present calculation neglects the contribution of the ITER ferromagnetic inserts. The active-coil field patterns shown here are computed from vacuum Biot–Savart fields and scaled by the relevant coil currents. Since the disambiguation algorithm uses a combined signature across multiple active-coil realisations, nulls in $|B \cdot \tau|$ for individual coils are not necessarily problematic provided they do not coincide across the full set of fields. Nevertheless, ferromagnetic-insert fringe fields (of order 10mT) may introduce systematic modelling errors in regions where the combined projected field amplitude, or more importantly the separation between neighbouring sensor signatures, is small. A quantitative assessment of this effect, including possible validation of effective ferromagnetic-insert parameters such as magnetic barycentre and saturation magnetisation, is left for future work.

In other future work, we anticipate verification of the technique through application to magnetic diagnostics on existing tokamaks, with the longer-term goal of reducing the commissioning time and diagnostic integration effort associated with ITER magnetic sensor systems.

\section*{Acknowledgements}
AI-assisted coding and editorial tools, including \lstinline|chatGPT|,  were used during development of the optimisation workflow and manuscript preparation. All mathematical formulation, verification, physical interpretation, and validation were performed by the authors. This work was performed as part of a visit to ITER by M. J. Hole as an ITER Scientist Fellow. The views and opinions expressed herein do not necessarily reflect those of the ITER Organization.

\printbibliography

@article{Hole_OMAHA_2009,
    title = {{A high resolution Mirnov array for the mega ampere spherical tokamak}},
    year = {2009},
    journal = {Review of Scientific Instruments},
    author = {Hole, M. J. and Appel, L. C. and Martin, R.},
    number = {12},
    volume = {80},
    doi = {10.1063/1.3272713},
    issn = {00346748}
}

@misc{BSmag_2015,
    title = {{BSmag Toolbox for Matlab}},
    year = {2015},
    author = {Qu{\'{e}}val, L.},
    publisher = {Dept. Elect. Eng., University of Applied Sciences D{\"{u}}sseldorf},
    address = {D{\"{u}}sseldorf}
}

@article{Appel_Hole_2005,
    title = {{Calibration of the high-frequency magnetic fluctuation diagnostic in plasma devices}},
    year = {2005},
    journal = {Review of Scientific Instruments},
    author = {Appel, L. C. and Hole, M. J.},
    number = {9},
    month = {9},
    volume = {76},
    doi = {10.1063/1.2009107},
    issn = {00346748}
}

@article{Abate_2022,
    title = {{Effective Area Measurements of Magnetic Pick-Up Coil Sensors for RFX-mod2}},
    year = {2022},
    journal = {Sensors},
    author = {Abate, Domenico and Cavazzana, Roberto},
    number = {24},
    month = {12},
    volume = {22},
    publisher = {MDPI},
    doi = {10.3390/s22249767},
    issn = {14248220},
    pmid = {36560135}
}

@article{Heeter_2000,
    title = {{Fast magnetic fluctuation diagnostics for Alfv{\'{e}}n eigenmode and magnetohydrodynamics studies at the Joint European Torus}},
    year = {2000},
    journal = {Review of Scientific Instruments},
    author = {Heeter, R. F. and Fasoli, A. F. and Ali-Arshad, S. and Moret, J. M.},
    number = {11},
    pages = {4092--4106},
    volume = {71},
    publisher = {American Institute of Physics Inc.},
    doi = {10.1063/1.1313797},
    issn = {00346748}
}

@article{Hole_2007b,
    title = {{Fourier decomposition of magnetic perturbations in toroidal plasmas using singular value decomposition}},
    year = {2007},
    journal = {Plasma Physics and Controlled Fusion},
    author = {Hole, M J and Appel, L C},
    number = {12},
    month = {12},
    pages = {1971--1988},
    volume = {49},
    publisher = {IOP Publishing},
    doi = {10.1088/0741-3335/49/12/002},
    issn = {0741-3335}
}

@book{Kay_1993,
    title = {{Fundamentals of statistical signal processing : volume 1 Estimation Theory}},
    year = {1993},
    author = {Kay, Steven M..},
    publisher = {Prentice-Hall PTR},
    isbn = {0133457117}
}

@book{Tarantola_2005,
    title = {{Inverse problem theory and methods for model parameter estimation}},
    year = {2005},
    author = {Tarantola, Albert},
    pages = {342},
    publisher = {Society for Industrial and Applied Mathematics},
    isbn = {0898715725}
}

@techreport{ITER_55.A0_2024,
    title = {{ITER Commissioning Document: 55.A0 - Magnetics Electronics and Software Technical Basis}},
    year = {2024},
    author = {Neto, A.},
    month = {2}
}

@article{Pustovitov_2001,
    title = {{Magnetic diagnostics: General principles and the problem of reconstruction of plasma current and pressure profiles in toroidal systems}},
    year = {2001},
    journal = {Nucl. Fusion},
    author = {Pustovitov, V D},
    pages = {721--730},
    volume = {41},
    doi = {10.1088/0029-5515/41/6/307}
}

@article{Moret_1998,
    title = {{Magnetic measurements on the TCV Tokamak}},
    year = {1998},
    journal = {Review of Scientific Instruments},
    author = {Moret, J. M. and Buhlmann, F. and Fasel, D. and Hofmann, F. and Tonetti, G.},
    number = {6},
    pages = {2333--2348},
    volume = {69},
    publisher = {American Institute of Physics Inc.},
    doi = {10.1063/1.1148940},
    issn = {00346748}
}

@article{Lao_2005,
    title = {{MHD equilibrium reconstruction in the DIII-D tokamak}},
    year = {2005},
    journal = {Fusion Science and Technology},
    author = {Lao, L. L. and St John, H. E. and Peng, Q. and Ferron, J. R. and Strait, E. J. and Taylor, T. S. and Meyer, W. H. and Zhang, C. and You, K. I.},
    number = {2},
    pages = {968--977},
    volume = {48},
    publisher = {American Nuclear Society},
    doi = {10.13182/FST48-968},
    issn = {15361055}
}

@book{Bishop_2009,
    title = {{Pattern recognition and machine learning}},
    year = {2009},
    author = {Bishop, Christopher M.},
    pages = {738},
    publisher = {Springer Science + Business Media},
    isbn = {0387310738},
    doi = {10.1117/1.2819119}
}

@book{Hutchinson_2002,
    title = {{Principles of Plasma Diagnostics}},
    year = {2002},
    author = {Hutchinson, I. H.},
    month = {7},
    publisher = {Cambridge University Press},
    isbn = {9780521803892},
    doi = {10.1017/CBO9780511613630}
}

@article{von_Nessi_2014,
    title = {{Recent developments in Bayesian inference of tokamak plasma equilibria and high-dimensional stochastic quadratures}},
    year = {2014},
    journal = {Plasma Physics and Controlled Fusion},
    author = {Von Nessi, G. T. and Hole, M. J.},
    number = {11},
    month = {11},
    pages = {114011},
    volume = {56},
    publisher = {Institute of Physics Publishing},
    doi = {10.1088/0741-3335/56/11/114011},
    issn = {13616587}
}

@article{Lao_1985,
    title = {{Reconstruction of current profile parameters and plasma shapes in tokamaks}},
    year = {1985},
    journal = {Nucl. Fusion},
    author = {Lao, L. L. and St John, H. and Stambaugh, R. D. and Kellman, A. G. and Pfeiffer, W.},
    pages = {1611--1622},
    volume = {25},
    doi = {10.1088/0029-5515/25/11/007}
}

@article{Kuhn_1955,
    title = {{The Hungarian method for the assignment problem}},
    year = {1955},
    journal = {Naval Research Logistics Quarterly},
    author = {Kuhn, H. W.},
    number = {1-2},
    month = {3},
    pages = {83--97},
    volume = {2},
    publisher = {Wiley},
    doi = {10.1002/nav.3800020109},
    issn = {0028-1441}
}

\end{document}